\documentclass[a4paper,fleqn]{cas-sc}

\usepackage[authoryear]{natbib}
\usepackage{longtable}
\usepackage{multirow}
\usepackage{siunitx}
\usepackage{pgfgantt,rotating}
\usepackage{tablefootnote}
\usepackage{subfloat}
\usepackage{multirow}
\usepackage{blindtext}
\usepackage{amsmath,caption,threeparttable,booktabs}
\usepackage{booktabs, caption}
\usepackage{amssymb,amsmath}
\usepackage{enumitem}
\usepackage{tikz}
\usepackage[T1]{fontenc}
\usepackage{wasysym}
\usepackage[font=small,labelfont=bf]{caption}
\usepackage{framed}
\usepackage{etoolbox}
\usepackage{array}
\usepackage{subfigure}
\usepackage[utf8]{inputenc}

\def\barr{\begin{array}}
	\def\earr{\end{array}}
\def\berr{\begin{eqnarray}}
\def\err{\end{eqnarray}}
\def\berrno{\begin{eqnarray*}}
	\def\errno{\end{eqnarray*}}
\renewcommand{\a}{\alpha}

\newcommand{\g}{\gamma}

\renewcommand{\d}{\delta}
\def\tsc#1{\csdef{#1}{\textsc{\lowercase{#1}}\xspace}}
\tsc{WGM}
\tsc{QE}

\begin{document}
	\let\WriteBookmarks\relax
	\def\floatpagepagefraction{1}
	\def\textpagefraction{.001}
	
	\shorttitle{V48, V49, V51, V55 and NV4}    
	
	\shortauthors{Shanti, Rukmini, Ravi, Dereje, Vineet, Safonova and Noah}  
	
	\title [mode = title]{Multi-wavelength photometric study of five contact binaries in the field of globular cluster M4}  
	
	
	
	%
	
	\author[1]{Shanti Priya Devarapalli}\corref{cor1}
	\ead{astroshanti@osmania.ac.in}
	
	\credit{Conceptualization (lead), writing-review and editing (equal)}
	
	\affiliation[a]{organization={Department of Astronomy},
		addressline={Osmania University}, 
		city={Hyderabad},
		postcode={500007}, 
		state={Telangana},
		country={India}}
	
	\author[1]{Rukmini Jagirdar}
	\ead{rukminiouastro@osmania.ac.in}
	\credit{Original Draft (lead), writing-review and editing (equal)}

	\author[1]{Ravi Raja Pothuneni}
	\ead{ravirajapothuneni@osmania.ac.in}
	\credit{Software (equal), formal analysis (equal), writing-review and editing (supporting)}

	\author[1]{Dereje Wakgari Amente}
	\credit{Software (equal), formal analysis (equal)}
	
	\author[2]{Vineet Thomas}
	\credit{Software (supporting), formal analysis (supporting), methodology (lead)}
	\affiliation[b]{organization={Northeastern University},
		addressline={}, 
		city={Boston},
		postcode={02115}, 
		state={MA},
		country={USA}}
	
	\author[3]{Margarita Safonova}
	\credit{Observed data collection (lead), review, and editing (supporting)}
	\affiliation[c]{organization={Indian Institute of Astrophysics},
		addressline={II Block, Koramangala,}, 
		city={Bengaluru},
		postcode={560 034}, 
		state={Karnataka},
		country={India}}
	
	\author[4]{Noah Brosch}
	\credit{Observed data collection (lead)}
	\affiliation[d]{organization={Tel Aviv University}, addressline={Tel Aviv}, country={Israel}}
	
	\cortext[1]{Shanti Priya Devarapalli}
	
	\fntext[1]{0000-0003-2672-8633}
	\fntext[2]{0000-0002-3007-5164}
	\fntext[3]{0000-0003-4938-0043}
	
	
	\begin{abstract}
		Binary stars are believed to be key determinants in understanding globular cluster evolution. In this paper, we present the Multi-band photometric analyses of five variables in the nearest galactic globular cluster M4, from the observations of CASE, M4 Core Project with HST for four variables (V48, V49, V51, and V55) and the data collected from T40 and C18 Telescopes of Wise Observatory for one variable (NV4). The light curves of five binaries are analyzed using the Wilson-Devinney method (WD) and their fundamental parameters have been derived. A period variation study was carried out using times of minima obtained from the literature for four binaries and the nature of the variation observed is discussed. The evolutionary state of systems is evaluated using M-R diagram, correlating with that of a few well-studied close binaries in other globular clusters. Based on the  data obtained from the Gaia DR3 database, a three-dimensional Vector-Point Diagrams (VPD) was built to evaluate the cluster membership of the variables, and two out of them (V49 and NV4) were found to be not cluster members.
	\end{abstract}
	
	
	\maketitle
	\section{Introduction}\label{intro}
	
	Messier 4 (M4/NGC 6121), is the closest globular cluster (GC) to the Sun and is observed to have a prominent photometric binary fraction in its core (\cite{kaluzny2013cluster}). The measured fraction is highest for a GC, reaching 15\% in the core region (\cite{milone2012acs}), thus making it interesting. Binaries play a key role in understanding the dynamical evolution of globular clusters and numerous eclipsing binaries have been discovered over the past decade. The study of eclipsing binaries in GCs not only tells about their impact on the dynamical evolution of GCs but also helps in understanding and determining the absolute dimensions of GC stars. Among the eclipsing binaries, W UMa binaries, in particular, are powerful tools to understand the parameters of GC as the components in these binaries are usually made up of two main sequence (MS) stars.
	
	Thus, the study of contact binaries gives an insight into stellar evolution theory through fundamental stellar parameters such as mass, radius, luminosity, and composition (\cite{yakut2005evolution}). These binaries have short periods (<1 d) and hence are appropriate to study the mechanism of binary mergers and may explain progenitors such as FK Comae Berenices, blue stragglers, etc. Only a few eclipsing binaries in globular clusters have been analyzed so far compared to field stars (\cite{hut1992binaries}, \cite{mateo1996photometric}, \cite{rucinski2000w}, \cite{milone2008photometric}, \cite{koccak2020photometric} and references within). To produce statistically reliable results large samples of binaries in the globular cluster must be surveyed. 
	
	In this paper, we present a detailed photometric study and orbital period investigation of four contact binaries, V48, V49, V51, and V55 (we use here identifiers given originally by \cite{kaluzny1997ccd}\footnote{For the current ID numbers of these variables, see Clement's Catalogue of Variable Stars in Galactic GCs at http://www.astro.utoronto.ca/~cclement/cat/C1620m264.}) from M4. The observations in the B and V bands were obtained from the CASE (The Cluster Ages Experiment) project, which was carried out between June 1995 and June 2009 (\cite{kaluzny2013cluster}), and the observations in the I band were obtained from the M4 Core Project using HST, which was carried out between October 2012 and September 2013 (\cite{nascimbeni2014m}). Apart from these, we also present the photometric analysis of a newly discovered variable NV4 \citep{safonova2016search}, for which the data was taken using 1-m and 0.46-m telescopes of the Wise Observatory, Israel. Our work draws motivation from the fact that very few contact binaries are studied and reported in globular clusters. The binary fraction is an essential parameter that can considerably affect the evolution of globular clusters and sufficient observational data of the same will help in testing the existing models, which are currently scarce. Here, we present the photometric analysis of five contact binaries in the M4 cluster, reporting the masses, luminosities, and radii of the components in order to understand their evolutionary state. 
	\begin{table}\footnotesize
		\centering
		\caption{Details of the variables under study (Identifiers, as given by \cite{kaluzny1997ccd}) and their rate of period variation.} 
		\begin{tabular}{ccccccc}
			\hline
			Variable Name&RA& Dec&P&V$_{mag}$&T$_{e,h}$& $\dot P$\\
			&\begin{math}(\alpha_{2000})\end{math}&\begin{math}(\alpha_{2000})\end{math}& (days)&&K&($\times$ 10 $^{-7}$ d/yr)\\
			\hline
			V48&16:23:36.81&-26:31:44.1&0.282694&16.8&6540&-0.656\\
			V49&16:23:34.35&-26:32:01.8&0.297444&17.1&4943&-0.375\\
			V51&16:23:33.28&-26:31:08.0&0.303683&17.1&5969&-4.051\\
			V55&16:23:45.78&-26:31:16.5&0.310703&16.7&6429&0.045\\
			\hline
			Variable Name&RA& Dec&P&I$_{mag}$&T$_{e,h}$& $\dot P$\\
			\hline
			NV4(V112)&16:25:09.48&-26:39:42.2&0.344440&15.3&5067&-\\
			\hline
		\end{tabular}
		\label{table1}
	\end{table}
	
	\section{Orbital Period Variations}\label{oc}
	
	Study of the orbital period changes in contact binaries can help in understanding the structure and evolutionary state of these systems, as period variations contribute directly in determining the mass transfer and angular momentum loss as predicted by TRO (Thermal Relaxation and Oscillation theory) and AML (Angular Momentum Loss) theory. All available times of light minima (ToMs) for primary eclipse have been collected for the observations between 1995 and 2009 (presented in Table~\ref{table2}). The ToMs are available only for four variables (V48, V49, V51, and V55) in the study as NV4 was observed for the first time. The Observed (O) minus Calculated (C) (O–C) values of the collected ToMs are computed and the corresponding (O–C) diagrams are plotted in Fig. \ref{figure1} (~\ref{figure1}a, \ref{figure1}b, \ref{figure1}c, and \ref{figure1}d). Based on the observed orbital period changes of the four contact binaries, we discuss their evolution status. 
	The following are the linear ephemerides to predict the primary minima $(Min. I$), where $E$ represents the epoch, or the number of cycles, used to calculate O-C values for the observed and calculated ToMs of the respective binaries. 
	\begin{eqnarray}
	&&Min.I_{48} = 2452767.7537+0.2827\times E\,,\nonumber\\
	&&Min.I_{49} = 2454969.8994+0.2974\times E\,,\nonumber\\
	&&Min.I_{51} = 2453108.8280+0.3037\times E\,,\nonumber\\
	&&Min.I_{55} = 2454629.6169+0.3107\times E\,. 		
	\end{eqnarray}

	\begin{eqnarray}
	V48&=&-2.540(\pm0.342)\times 10^{-11}\times E^2 +6.347(\pm 0.366)\times10^{-7}\times E-2.685(\pm0.356)\times10^{-4} ,\nonumber\\ 
	V49&=&-1.563(\pm0.451)\times10^{-11}\times E^2-4.080(\pm0.332)  \times10^{-7}\times E+3.430(\pm0.105)\times10^{-5}\nonumber\\&& -0.0011(\pm0.0005)\times sin(5.616(\pm0.086)\times10^{-4}+3.767(\pm0.668), \nonumber\\
	V51&=&-1.684(\pm0.302)\times 10^{-10}\times E^2 -8.483(\pm2.488)\times10^{-7}\times E+7.989(\pm2.264) \times10^{-4}, \nonumber\\
	V55&=&1.902(\pm0.421)\times 10^{-12}\times E^2 +5.279 (\pm0.0.750)\times 10^{-7}\times E+2.223(\pm0.327) \times 10^{-3}
	\end{eqnarray}
		The (O-C) diagrams show that V48, V49 \& V51 show a downward parabolic trend, suggesting a secular decrease in their period,  while V55 shows an upward parabolic trend, suggesting a secular increase in its period. The continuous period decrease/increase rates were derived using the following quadratic fit, obtained from their respective (O-C) plots, and the derived $\dot P$ for the objects in the study are listed in Table~\ref{table1}. \\
	In the case of V49, we have observed cyclic variation which is plotted by a sine fit on the residuals obtained from the quadratic fit. In the Fig. \ref{figure1}b, the quadratic+sine fit is represented by a thick line and the quadratic fit by a dashed line in the top panel and the residuals fit of quadratic+sine fit in the bottom panel.
	\begin{longtable}{llll|llll}
		\caption{Times of Light Minima for primary eclipse of V48, V49, V51 and V55.} \label{table2}\\
		\hline ToM&Epoch&(O-C)&Res&ToM&Epoch&(O-C)&Res\\ \hline
		\hline (2450000+)(d)&&(d)&(d)&(2450000+)(d)&&(d)&(d)\\ \hline 
		\endfirsthead
		\multicolumn{3}{c}%
		{{\bfseries \tablename\ \thetable{} -- continued from previous page}}\\
		\hline ToM&Epoch&(O-C)&Res&ToM&Epoch&(O-C)&Res\\ \hline 
		\hline (2450000+)(d)&&(d)&(d)&(2450000+)(d)&&(d)&(d)\\ \hline 
		\endhead
		\hline \multicolumn{7}{r}{{contd ..}} \\ \hline
		\endfoot
		\endlastfoot
		\hline
		\textbf{V48}&&&&\textbf{V48}&&&\\
		\hline
		49869.5678&-10252&-0.0070&0.0024&49873.8057&-10237&-0.0095&-0.0001\\
		49869.8453&-10251&-0.0122&-0.0028&52765.7743&-7&-0.0005&-0.0003\\
		49870.6952&-10248&-0.0104&-0.0009&52767.7537&0&0.0000&0.0003\\
		49871.8283&-10244&-0.0081&0.0014&54969.6604&7789&0.0031&0.0000\\
		\hline
		\hline
		\textbf{V49}&&&&\textbf{V49}&&\\
		\hline
		49869.6291&-17147&0.0020&0.0000&53859.8379&-3732&-0.0005&-0.0008\\
		52033.8329&-9871&0.0032&-0.0003&53911.5944&-3558&0.0008&0.0005\\
		52402.6643&-8631&0.0041&0.0007&53912.4865&-3555&0.0005&0.0003\\
		52763.7597&-7417&0.0024&-0.0002&54656.6919&-1053&0.0010&0.0006\\
		53107.9018&-6260&0.0018&0.0001&54969.6031&-1&0.0011&0.0005\\
		53108.7937&-6257&0.0014&-0.0003&54969.8994&0&0.0000&-0.0006\\
		53202.4891&-5942&0.0019&0.0005\\
		\hline
		\hline
		\textbf{V51}&&&&\textbf{V51}&&&\\
		\hline
		49869.4331&-10667&-0.0083&0.0010&53108.8280&0&0.0000&-0.0008\\
		49869.7351&-10666&-0.0100&-0.0007&53154.6848&151&0.0007&0.0000\\
		52033.7905&-3540&0.0003&-0.0014&53205.7029&319&0.0000&-0.0005\\
		52402.7672&-2325&0.0022&0.0003&53859.8358&2473&-0.0003&0.0021\\
		52763.8469&-1136&0.0028&0.0012&54656.6919&5097&-0.0084&-0.0005\\
		53107.9151&-3&-0.0019&-0.0027&54969.7867&6128&-0.0107&0.0000\\
		52767.7948&-1123&0.0028&0.0013\\
		\hline
		\hline
		\textbf{V55}&&&&\textbf{V55}&&&\\
		\hline
		50966.7355&-11789&-0.0037&0.0000&53859.6958&-2478&0.0009&0.0000\\
		53107.7927&-4898&-0.0009&-0.0006&53860.6273&-2475&0.0003&-0.0006\\
		53146.6316&-4773&0.0001&0.0004&53911.5843&-2311&0.0020&0.0010\\
		53154.7092&-4747&-0.0006&-0.0003&54629.6169&0&0.0000&-0.0022\\
		53202.557&-4593&-0.0010&-0.0009&54656.6504&87&0.0023&0.0001\\
		53205.6657&-4583&0.0006&0.0008&54969.8398&1095&0.0031&0.0003\\
		53858.7648&-2481&0.0020&0.0011\\
		\hline
	\end{longtable}
	\begin{figure*}[!h]
		\centering
		\includegraphics[width=0.9\textwidth]{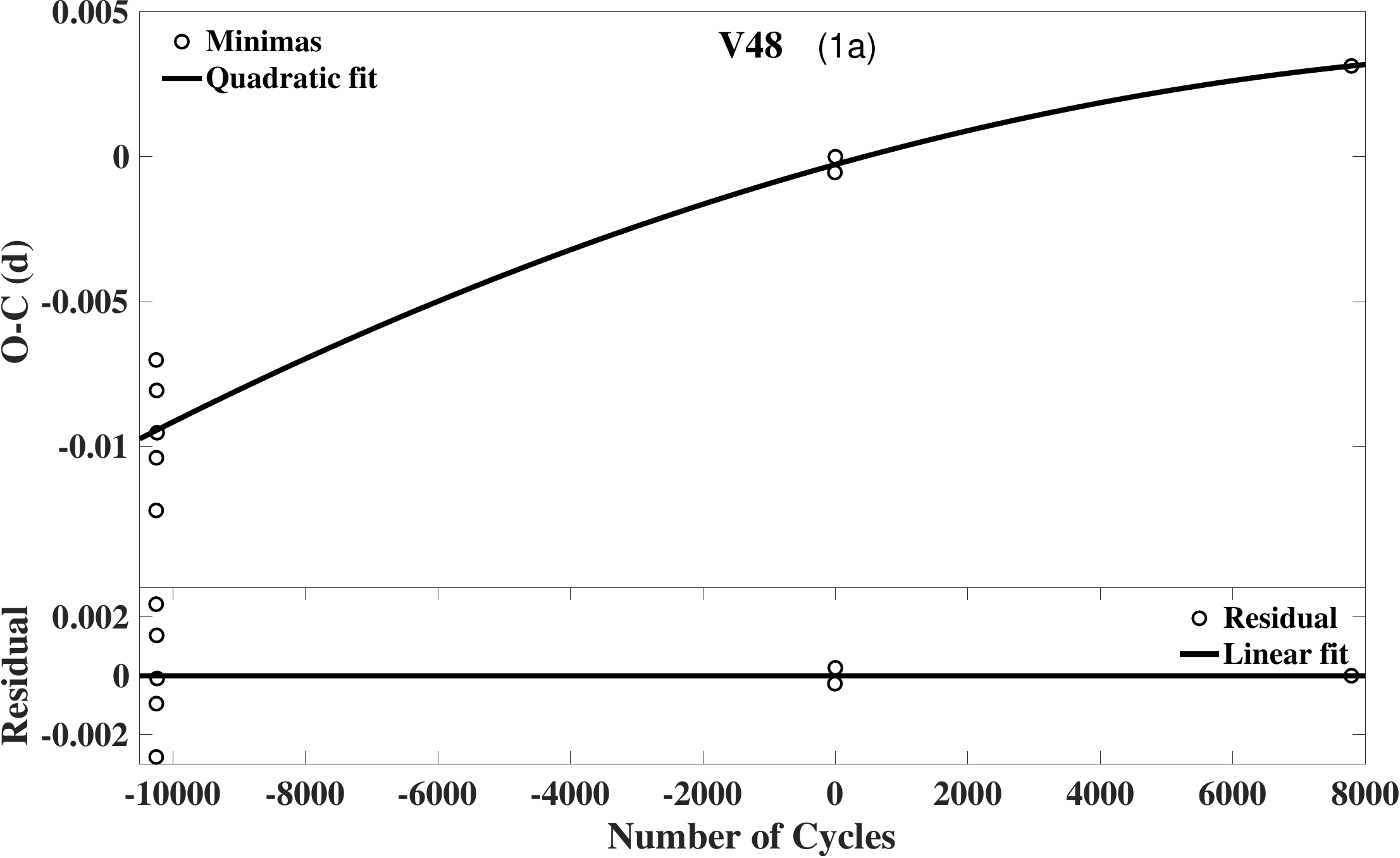}\\
		\includegraphics[width=0.85\textwidth]{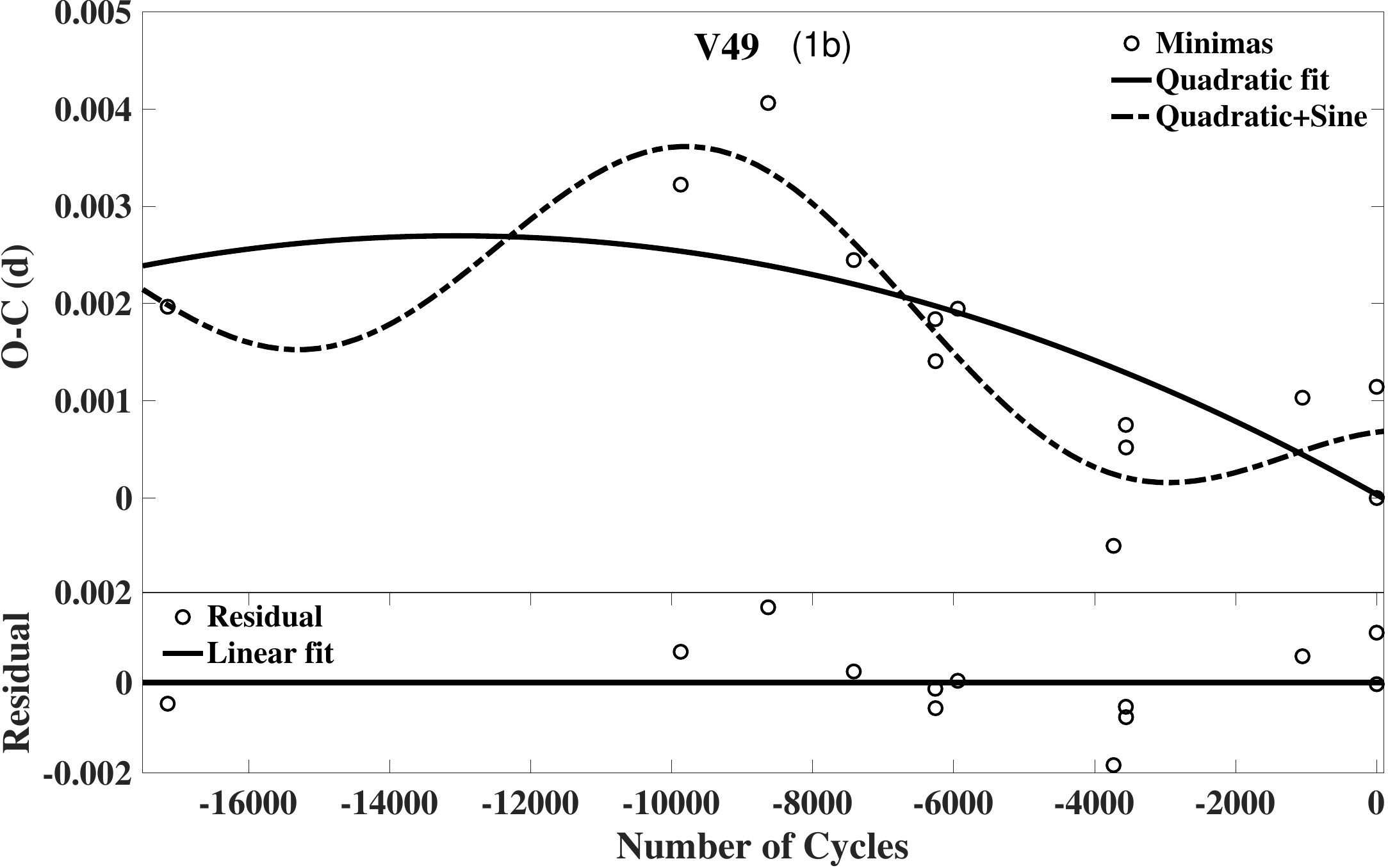}\\
	\end{figure*}
	\begin{figure}
		\centering
		\includegraphics[width=0.85\textwidth]{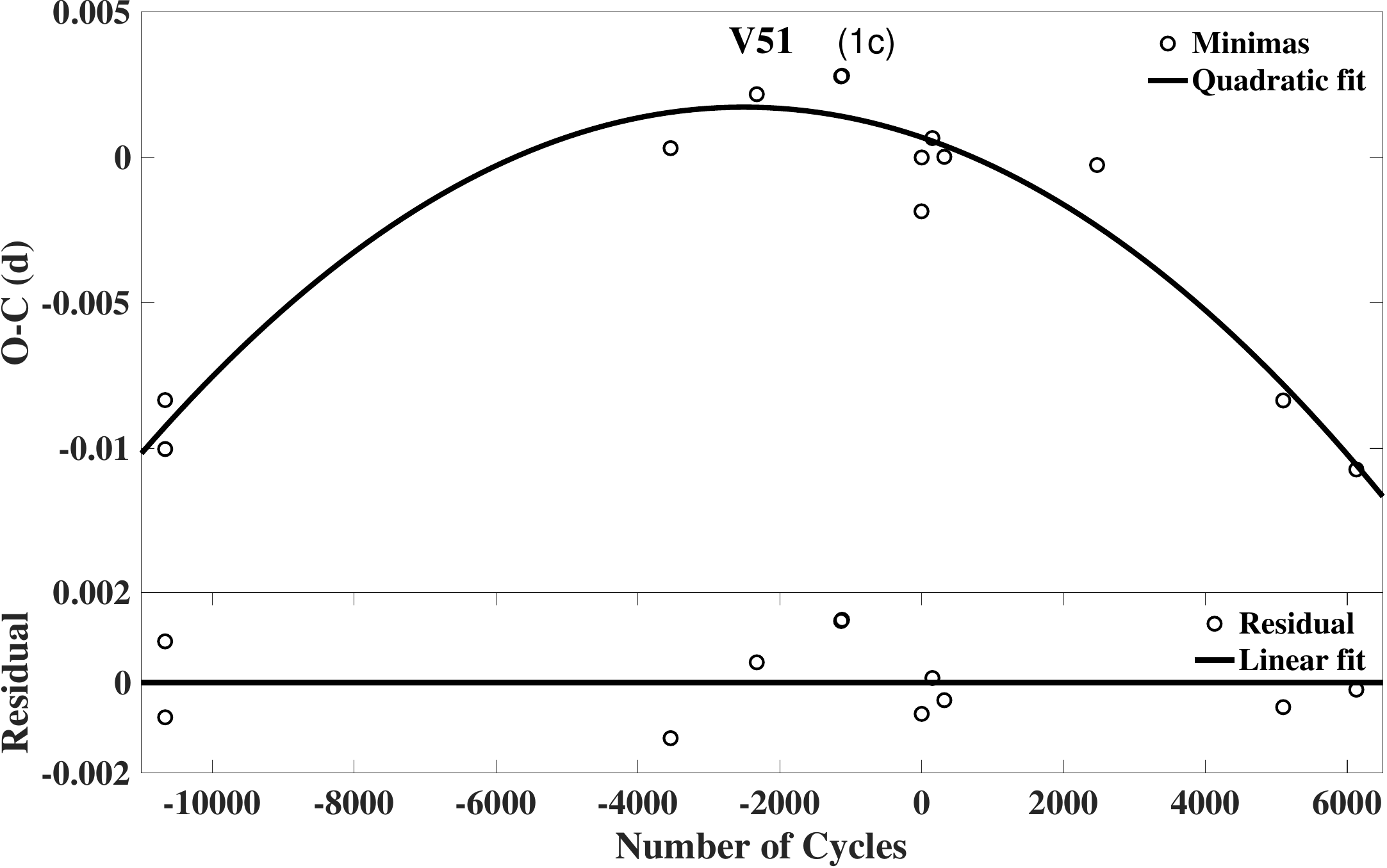}\\
		\includegraphics[width=0.9\textwidth]{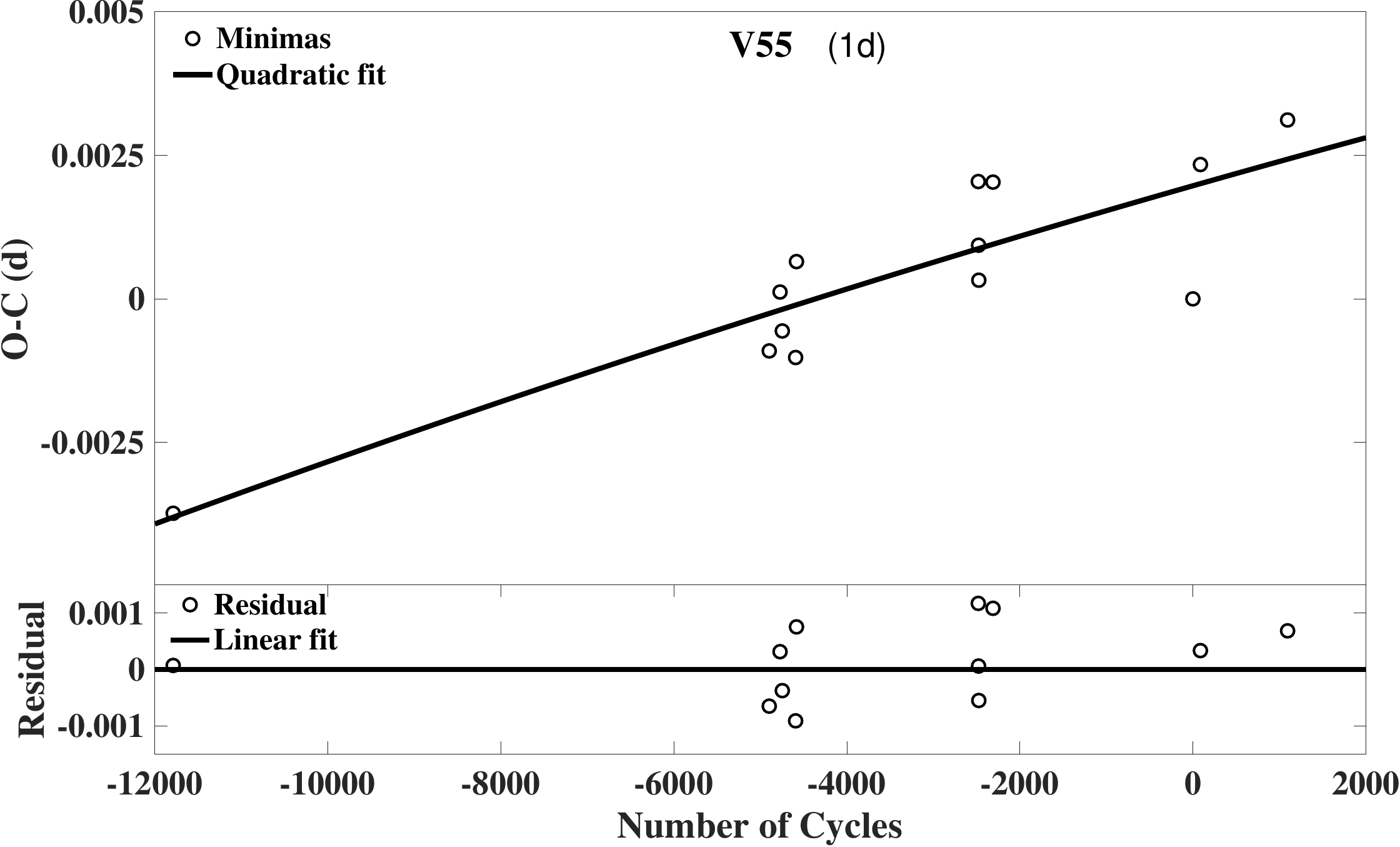}\\
		\caption{The (O-C) Curves of four variables and their best quadratic fitting.}
		\label{figure1}
	\end{figure} 
	\begin{table}[]\footnotesize
		\centering
		\caption{Photometric solutions of five variables under study in the different pass bands.}
		\begin{tabular}{cc|cccc}
			\hline
			Photometric& {\bf NV4(V112)}&& {\bf V48} &\\
			Elements&I, R&B, V&&I\\
			\hline
			T$_{e,c}$(K)&4318$\pm$16&6568$\pm$14&&6420$\pm$18\\
			q&1.216$\pm$0.015&0.152$\pm$0.001&&0.152$\pm$0.001\\
			i$^{o}$&56.58$\pm$0.14&80.04$\pm$0.43&&81.67$\pm$0.42\\
			q$_{min}$, $\Omega_{in}$, $\Omega_{out}$&1.3, 4.2233, 3.6571&0.16, 2.1563, 2.0471&&0.17, 2.1299, 2.0269\\
			$\Omega_h$=$\Omega_c$&4.0785$\pm$0.0118&2.0906$\pm$0.0047&&2.0677$\pm$0.0110\\
			f\%&26&51&&61\\
			\begin{math}\frac{L_{h}}{L_{h}+L_{c}}\end{math},\begin{math}\frac{L_{c}}{L_{h}+L_{c}}\end{math}&0.49,0.51; 0.57, 0.43&0.83,0.17; 0.83,0.17&&0.84; 0.16\\
			x$_{h}$&0.6&0.6&&0.6\\ 
			x$_{c}$&0.6&0.6&&0.6\\
			A$_{h,c}$&0.5&1&&1\\
			g$_{h,c}$&0.32&0.5&&0.5\\
			w(O-C)$^2$&0.26&0.48&&0.57\\
			\hline
			Spot Parameters&&&&\\
			Star&2&1, 2&&1, 2\\
			Co-latitude ($^{o}$)&78&90, 80&&90, 80\\
			Longitude ($^{o}$)&90&85, 125&&85, 125\\
			Radius ($^{o}$)&20&15, 15&&15, 15\\
			Temperature factor&1.25&1.03, 0.87&&1.03, 0.87\\
			\hline
		\end{tabular}
		\begin{tabular}{cccc|c}
			\hline
			Photometric& &{\bf V51} && {\bf V55} \\
			Elements&B, V&&I&B, V\\
			\hline
			T$_{e,c}$K)&5920$\pm$18&&5620$\pm$21&5785$\pm$16\\
			q&1.251$\pm$0.041&&1.274$\pm$0.024&1.247$\pm$0.002\\
			i$^{o}$&67.04$\pm$0.20&&66.46$\pm$0.38&64.51$\pm$0.14\\
			q$_{min}$, $\Omega_{in}$,$\Omega_{out}$&1.3, 4.2233, 3.6571&&1.3, 4.2233, 3.6571&1.3, 4.2233, 3.6571\\
			$\Omega_h$=$\Omega_c$&4.0784$\pm$0.0608&&4.0664$\pm$0.0427&4.1174$\pm$0.0051\\
			f\%&26&&28&19\\
			\begin{math}\frac{L_{h}}{L_{h}+L_{c}}\end{math},\begin{math}\frac{L_{c}}{L_{h}+L_{c}}\end{math}&0.42, 0.58; 0.44, 0.56&&0.44, 0.56&0.48, 0.52\\
			x$_{h}$&0.6&&0.6&0.6\\
			x$_{c}$&0.6&&0.6&0.6\\
			A$_{h,c}$&0.5&&0.5&1\\
			g$_{h,c}$&0.32&&0.32&0.5\\
			w(O-C)$^2$&0.25&&0.28&0.18\\
			\hline
			Spot Parameters&&&&\\
			Star&1&&1&1\\
			Co-latitude ($^{o}$)&70&&71&75\\
			Longitude ($^{o}$)&84&&85&72\\
			Radius ($^{o}$)&25&&34&20\\
			Temperature factor&0.96&&0.99&0.98\\
			\hline
		\end{tabular}
		\begin{tabular}{cccc}
			\hline		
			Photometric& &{\bf V49} &\\
			Elements&B, V&&I\\
			\hline
			T$_{e,c}$(K)&4676$\pm$6&&4658$\pm$11\\		q&1.184$\pm$0.020&&1.144$\pm$0.018\\
			i$^{o}$&86.20$\pm$0.22&&84.20$\pm$0.24\\
			q$_{min}$, $\Omega_{in}$, $\Omega_{out}$&1.2, 4.0684, 3.5088&&1.2, 4.0684, 3.5088\\
			$\Omega_h$=$\Omega_c$&3.7764$\pm$0.0262&&3.8264$\pm$0.0241\\
			f\%&52&&43\\
			\begin{math}\frac{L_{h}}{L_{h}+L_{c}}\end{math},\begin{math}\frac{L_{c}}{L_{h}+L_{c}}\end{math}&0.56, 0.44; 0.52,0.48&&0.47, 0.53\\
			x$_{h}$&0.6&&0.6\\ 
			x$_{c}$&0.6&&0.6\\
			A$_{h,c}$&0.5&&0.5\\
			g$_{h,c}$&0.32&&0.32\\
			w(O-C)$^2$&0.13&&0.18\\
			\hline
		\end{tabular}	\label{table3}
	\end{table}
	\begin{figure*}[!h]
		\centering
		\includegraphics[width=0.8\textwidth]{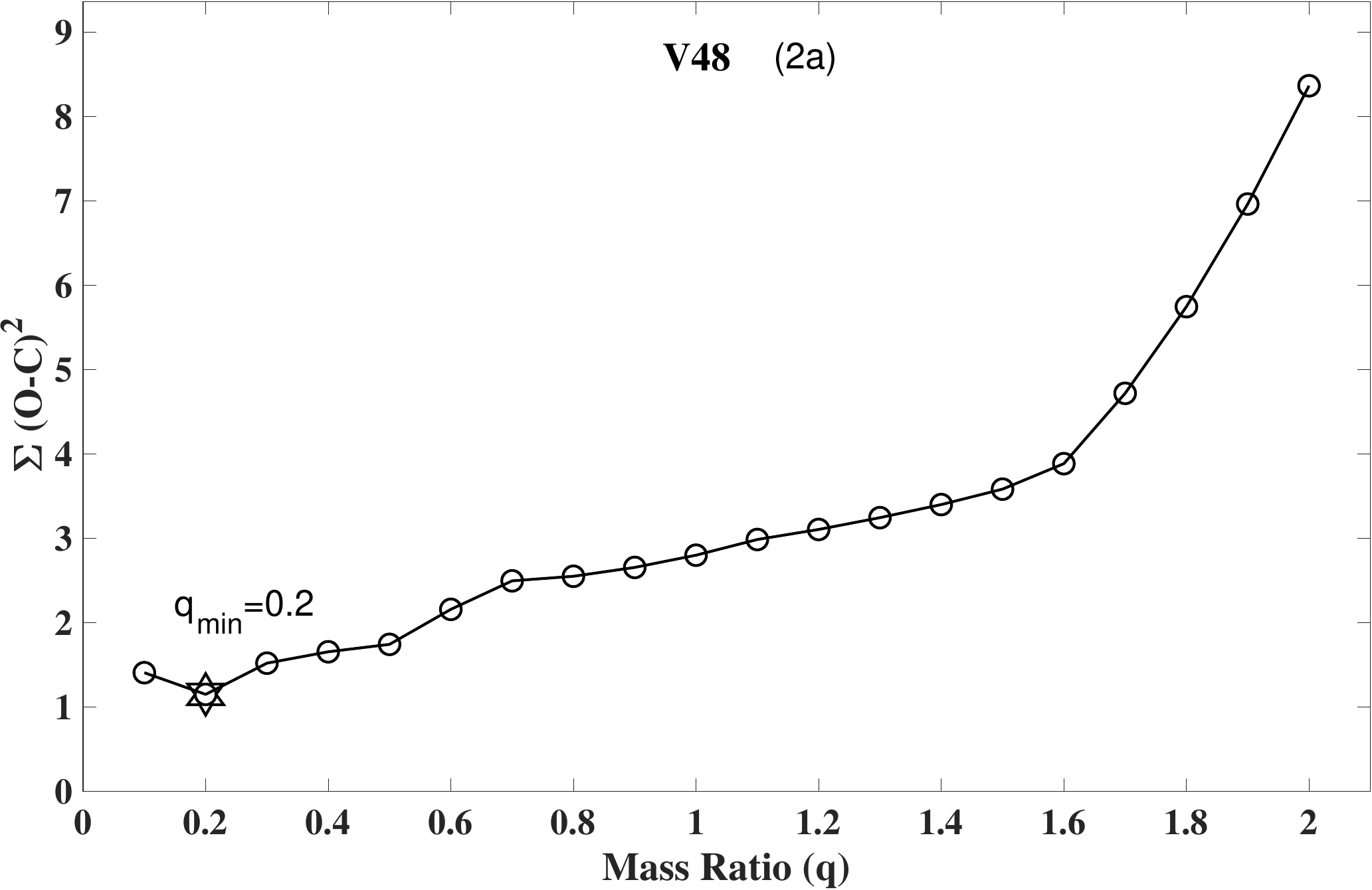}\\
		\includegraphics[width=0.8\textwidth]{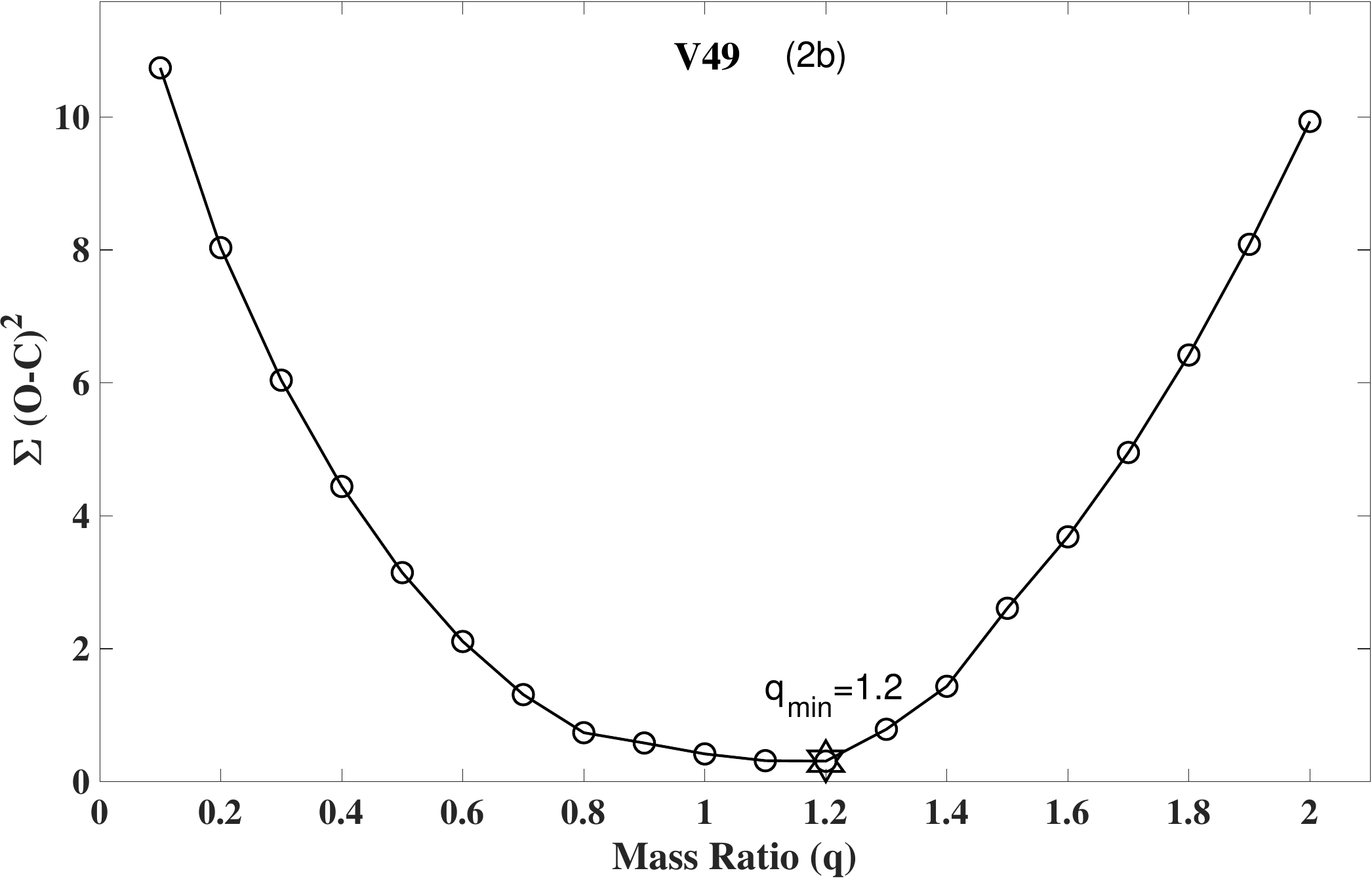}\\
	\end{figure*}
	\begin{figure*}[!h]
		\centering 
		\includegraphics[width=0.8\textwidth]{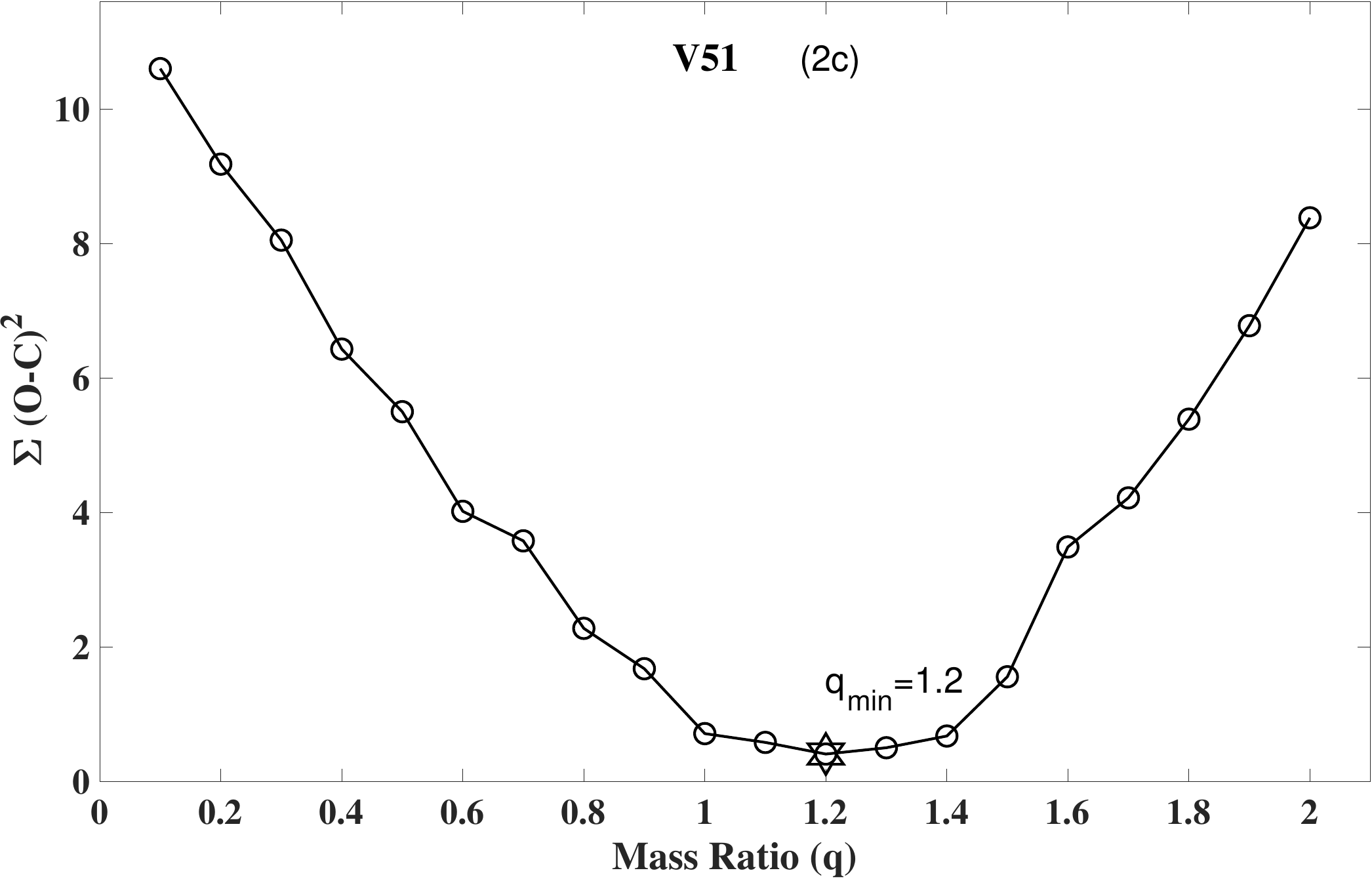}\\
		\includegraphics[width=0.8\textwidth]{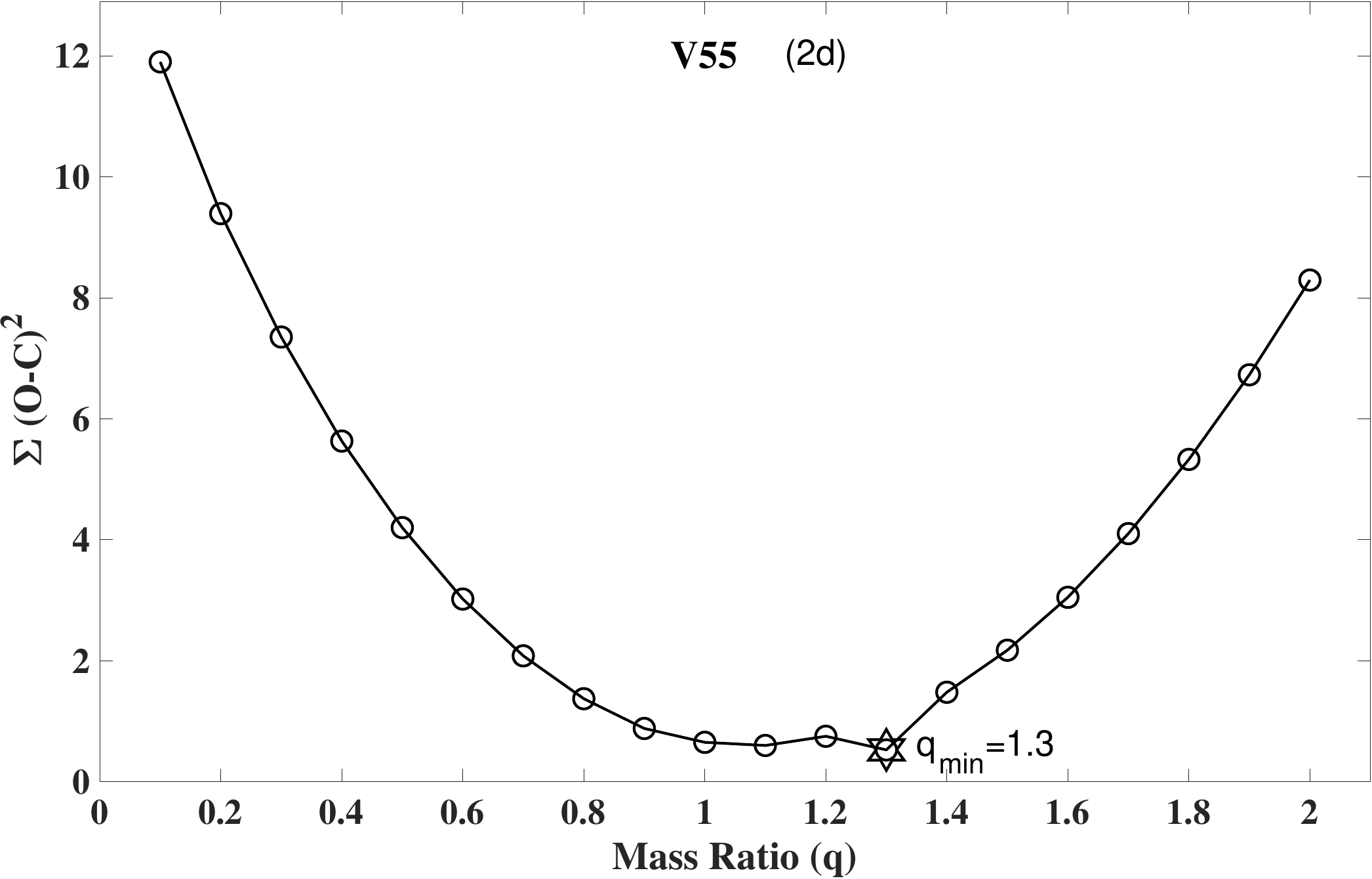}\\
	\end{figure*}
	\begin{figure*}[!h]
		\centering 
		\includegraphics[width=0.8\textwidth]{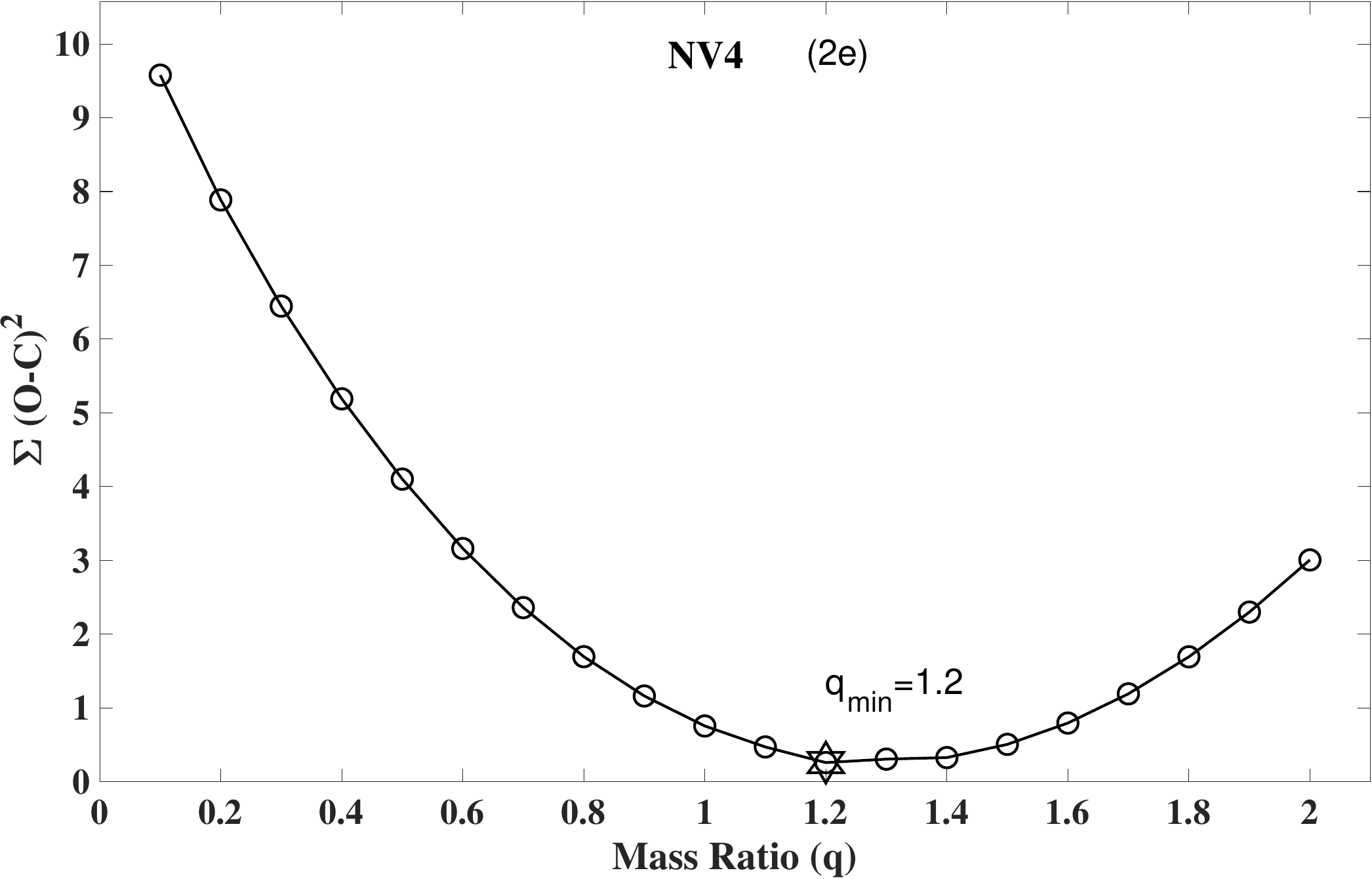}
		\caption{The q-search plot featuring minimum $\Sigma(O-C)^2$ obtained at q$_{min}$.}
		\label{figure2}
	\end{figure*}   
	\begin{figure*}[!h]
		\centering 
		\includegraphics[width=0.8\textwidth]{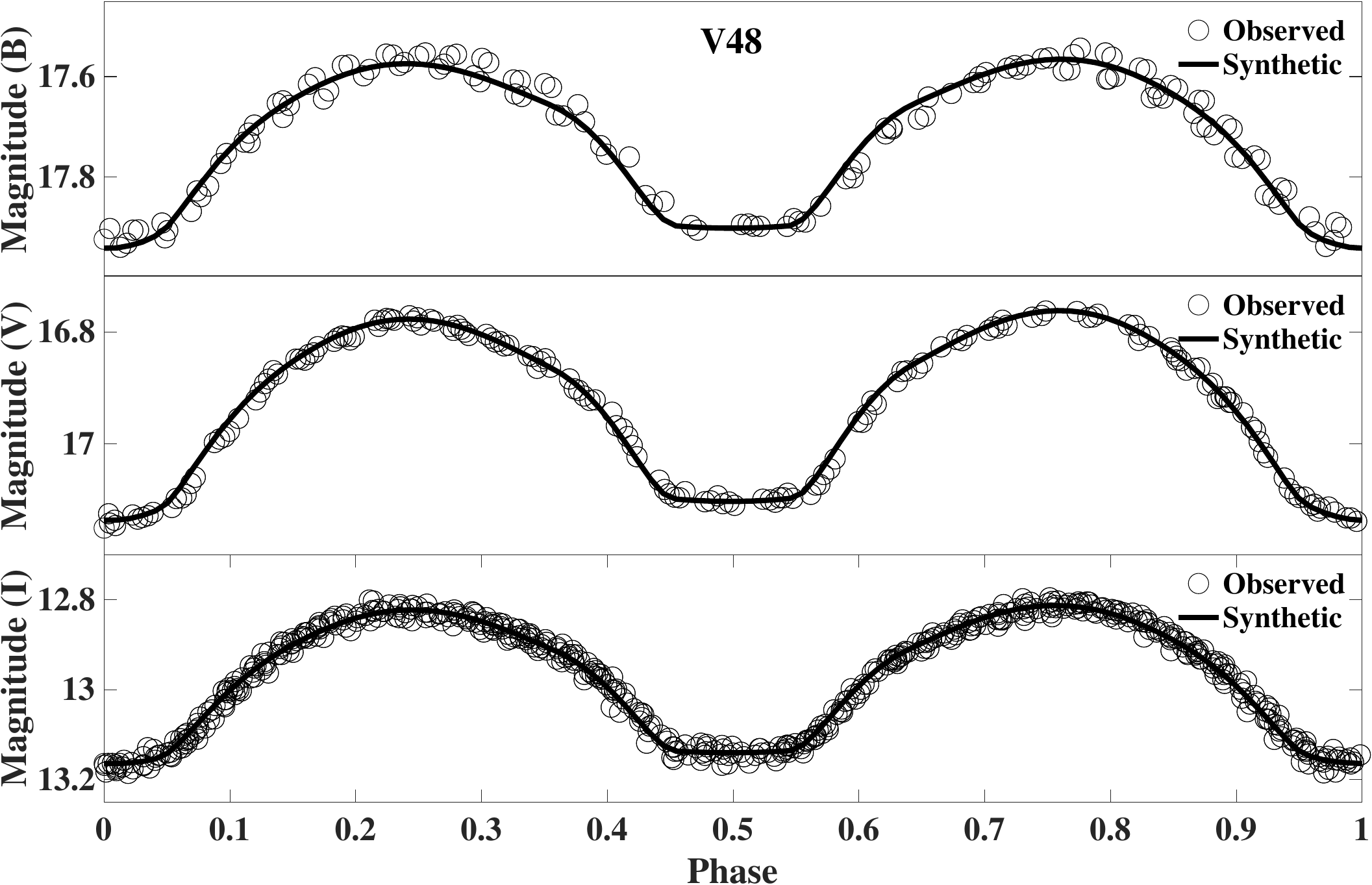}\\
		\includegraphics[width=0.8\textwidth]{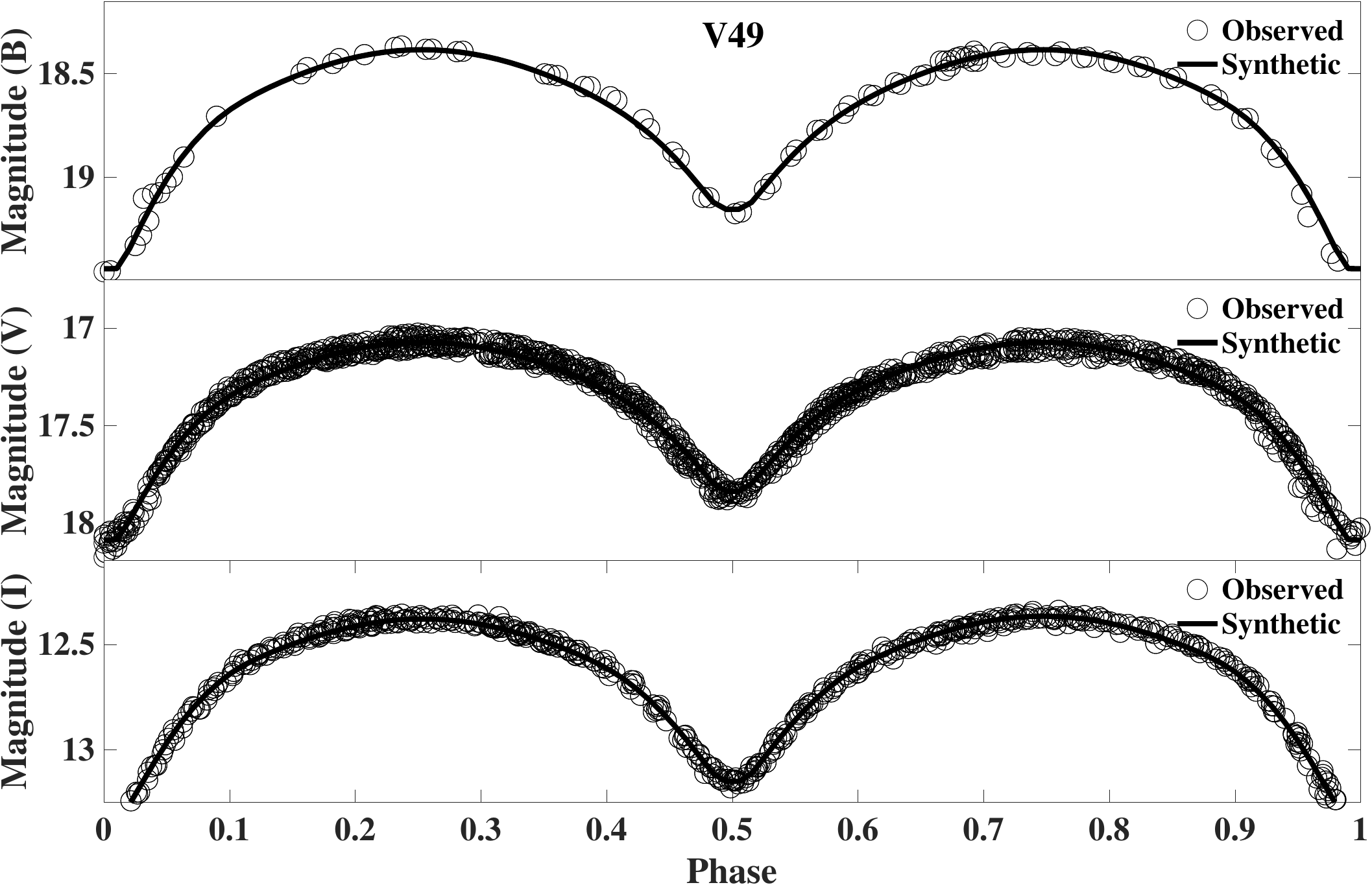}\\
	\end{figure*}
	\begin{figure*}[!h]
 		\centering 
		\includegraphics[width=0.8\textwidth]{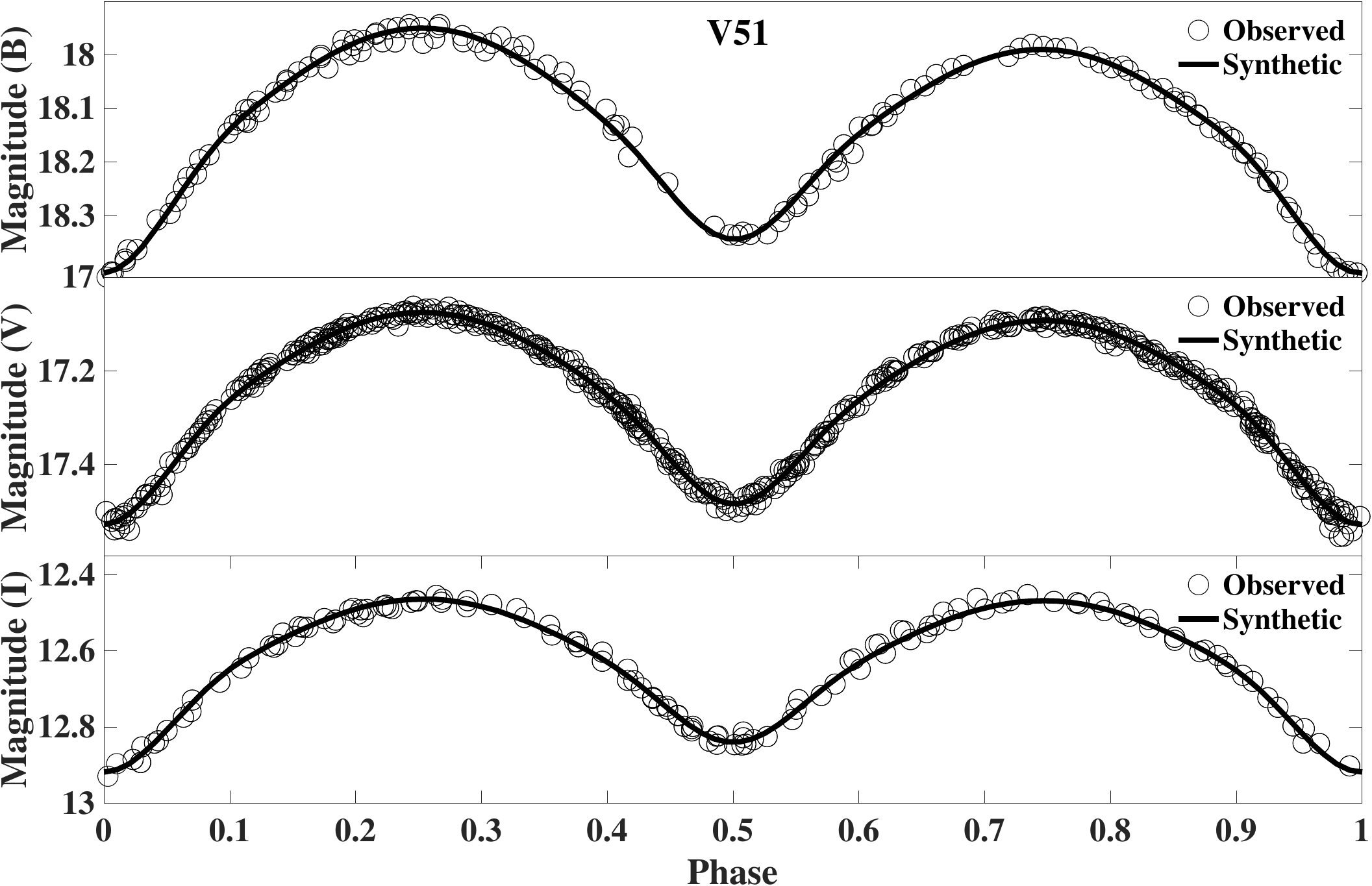}\\
		\includegraphics[width=0.8\textwidth]{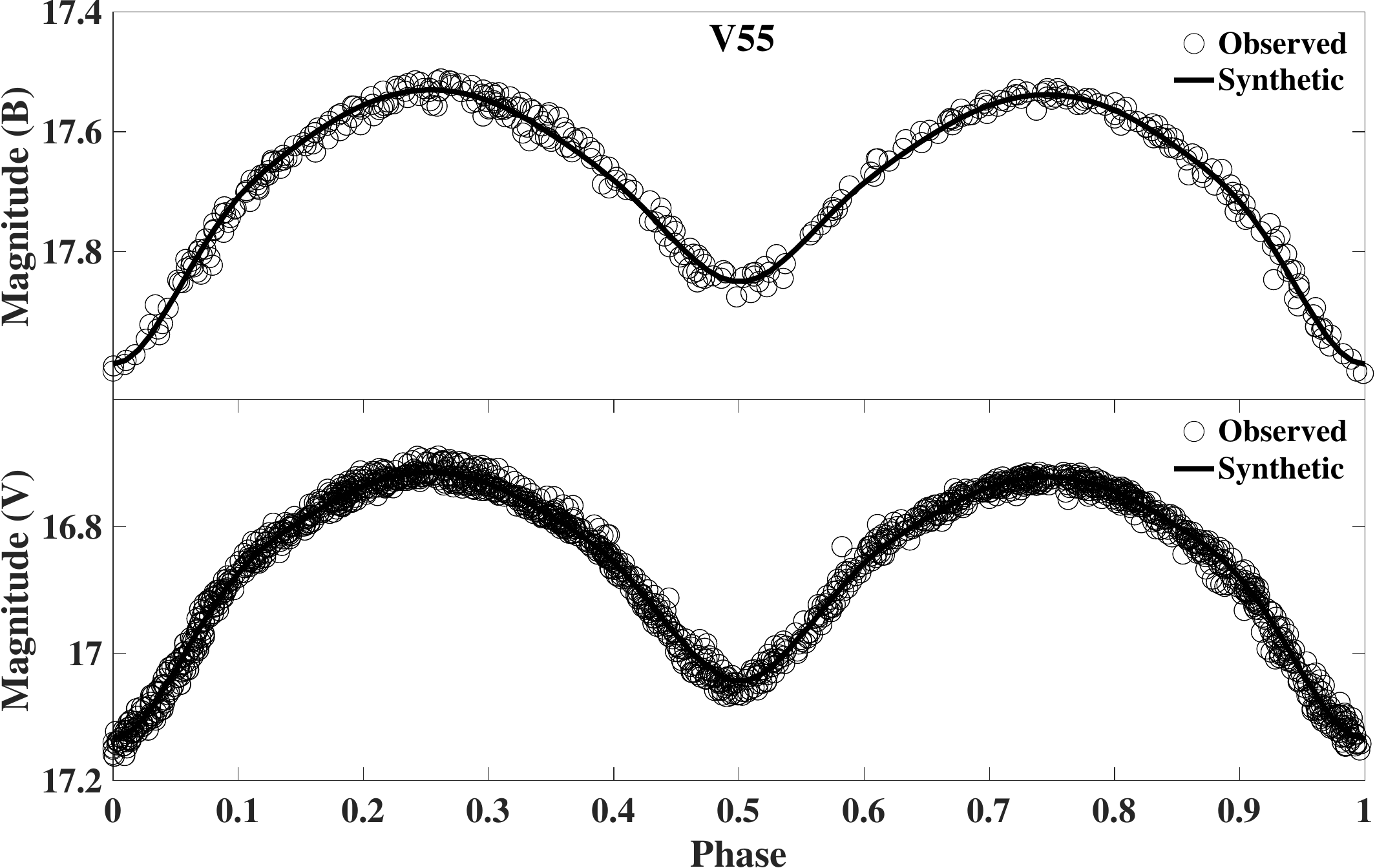}\\
	\end{figure*} 
	\begin{figure*}[!h]
		\centering 
		\includegraphics[width=0.8\textwidth]{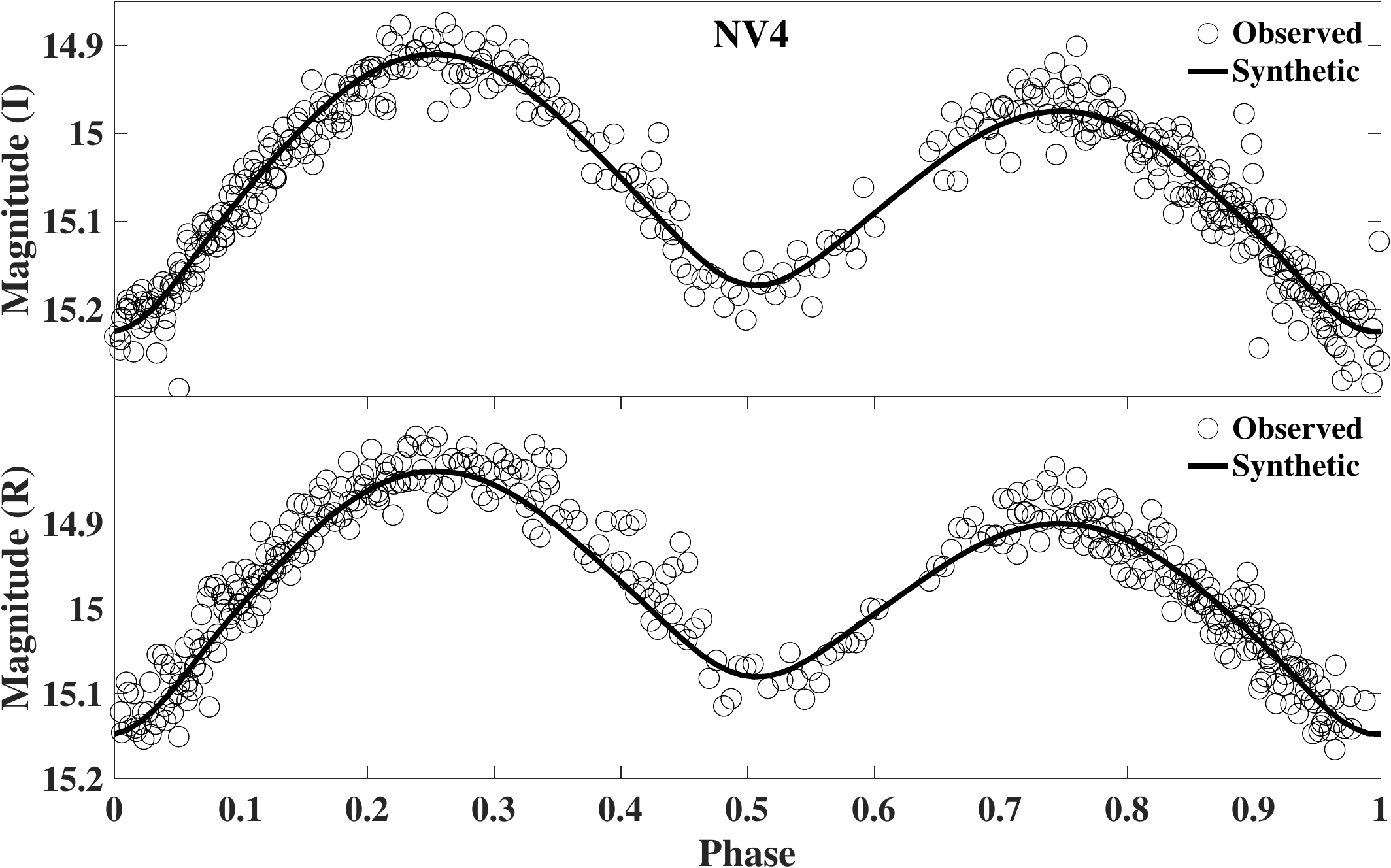}
		\caption{The best-fit light curves of the variables V48, V49, V51, V55 and NV4.}
		\label{figure3}
	\end{figure*}   
	\onecolumn
	\section{Photometric Observations and Analysis}
	
	The observational data for variables V48, V49, V51, and V55 in the B and V bands were taken from CASE (The Cluster Ages Experiment) conducted during June 1995 - June 2009 (\cite{kaluzny2013cluster}), and the I band data was taken from M4 Core Project with HST, conducted during October 2012 - September 2013 (\cite{milone2014m,nascimbeni2014m}). The 1-m (T40) telescope and 0.46-m Centurion 18 (C18) telescope of the Wise Observatory, Tel Aviv, Israel, were used to collect data for variable NV4 (V112 in Clement's Catalogue), respectively, over the periods of 06 April and 07 July 2011 and on six consecutive nights from 28 May to 02 June 2011. Further details of the observations and equipment used are given in \cite{safonova2016search}. A semi-automated pipeline was used to extract light curve from the collected images for each of the bands. The photometric solutions were obtained using the Wilson–Devinney (W-D) 2003 program (\cite{wilson1971realization,van2003stellar}). Orbital periods of the variables were determined using the Lomb-Scargle periodogram \cite{scargle1982studies,zechmeister2009generalised}, and are found to be similar to the values obtained by \cite{Kaluzny2013new}. The effective temperatures of the primary components were fixed using \cite{pecaut2013intrinsic} after correcting the observed B-V values of \cite{Kaluzny2013new} using the latest reddening value given by \cite{richer2004concerning} as E(B-V)={A$_V$}/{R$_V$}={1.33}/{3.8}=0.35. The gravity darkening coefficients (g$_1$, g$_2$), limb darkening coefficients (x$_1$, $x_2$), and bolometric albedos ($A_1$, $A_2$) of the components were adopted suitably with respect to their convective and radiative envelopes as suggested by \cite{lucy1967structure,rucinski1969proximity} and \cite{van1993new}. They were taken as fixed parameters throughout the analysis. The temperature of the secondary component T$_2$, orbital inclination ($i$), the mass ratio ($q$), the dimensionless potentials of the components $\Omega_1$=$\Omega_2$ and the bandpass luminosity of the primary star L$_1$ were taken as adjustable parameters. The contact configuration, i.e. mode 3 option of the WD Program was used to fit the light curve and determine the parameters after over-ruling the mode 2 option, as the solutions diverged. The method adopted for modeling light curves is briefly discussed in \cite{pothuneni2023first}, \citealt{joshi2016photometric}, and \cite{priya2013photometric}. As no spectroscopic value of $q$ was available, the $q$-search method for values of q from 0.05 to 5 at an interval of 0.1 was used to determine its mass ratio. Fig. \ref{figure2} (V48, V49, V51, V55 and NV4) displays the sum of squares of residuals $\Sigma(O-C)^2$ for each assumed value of q. The best-fit photometric solutions obtained are listed in Table \ref{table3}. The corresponding light curves, for all the variables in different filters, are plotted in Fig. \ref{figure3} (V48, V49, V51, V55 and NV4), and a good agreement were observed between the observed and synthetic light curves. Using three-dimensional correlations by \cite{gazeas2009physical}, we have determined the absolute physical parameters of these variables (Table~\ref{table4}). They were found to be showing similar properties to that of other contact binaries in GCs.\\ 
	\begin{figure}[h!]
		\centering
		\includegraphics[width=0.9\textwidth]{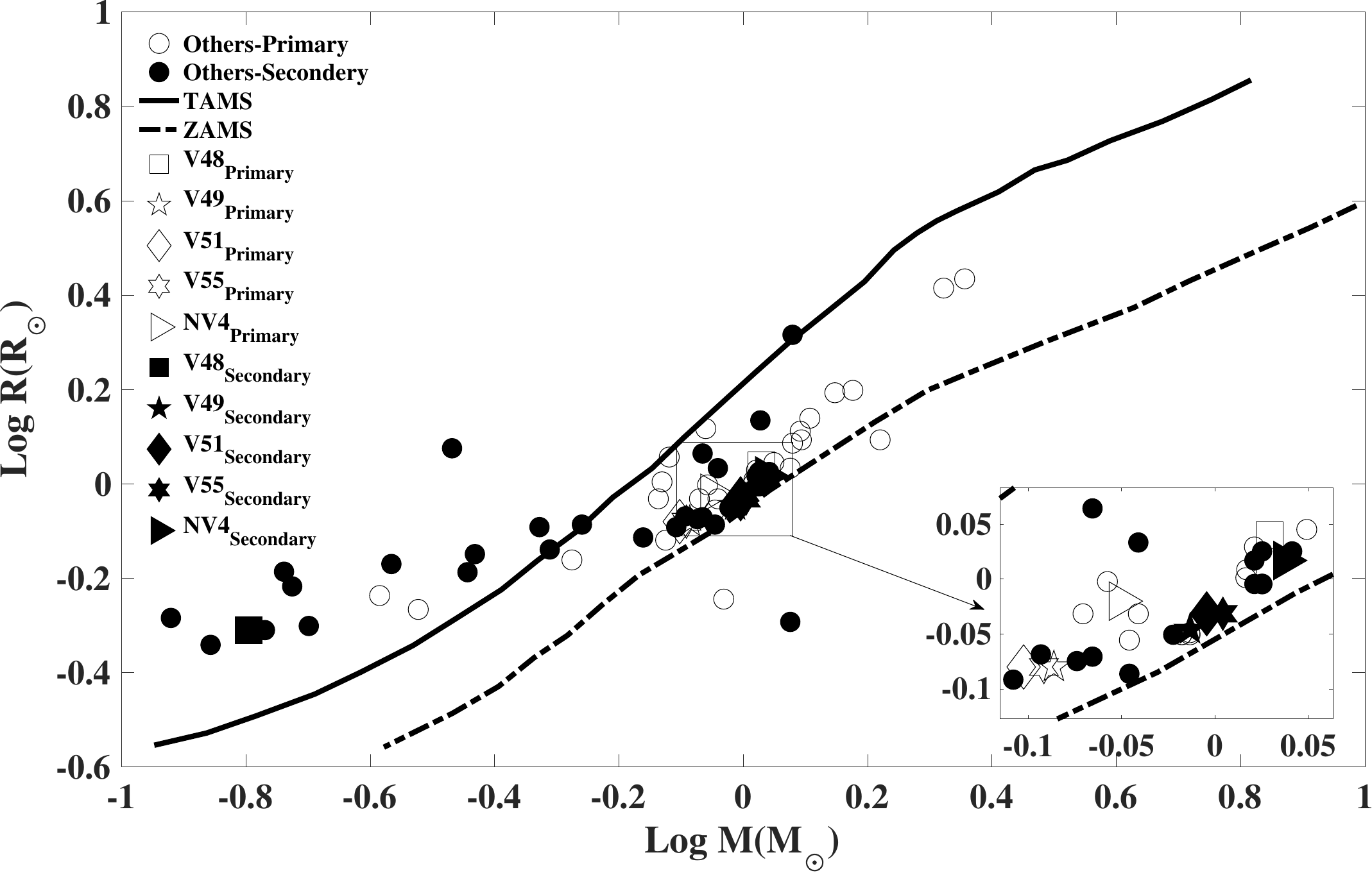}
		\caption{The Mass–Radius diagram. The solid line represents the zero-age main sequence (ZAMS), and the dashed line represents the terminal-age main sequence (TAMS). The lines are adopted from \cite{ma2022photometric}. The filled circles represnt components of contact binaries in clusters listed in Table \ref{table5}.}
		\label{figure4}
	\end{figure}
	\begin{figure}[!h]
		\centering
		\includegraphics[width=0.9\textwidth]{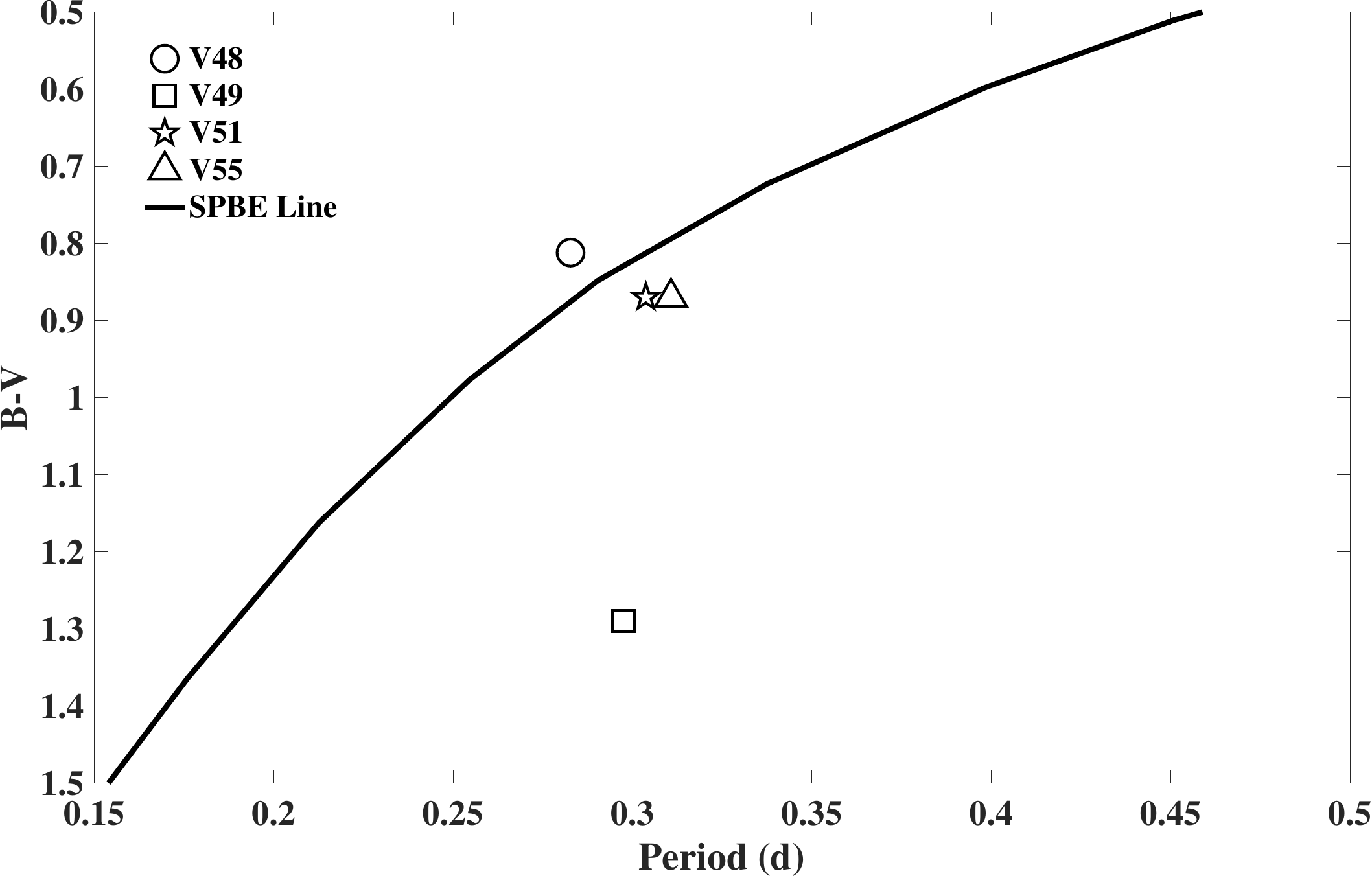}
		\caption{Period-color diagram. The solid line represents the SPBE theoretical line adopted from \cite{rucinski1997absolute}.}
		\label{figure5}
	\end{figure}
	\begin{table}[!b]\footnotesize
		\centering
		\caption{Derived absolute parameters of the variables.}
		\begin{tabular}
			{cccccccccc}
			\hline
			Variable&M$_1$&M$_2$&R$_1$&R$_2$&L$_1$&L$_2$&J$_{orb}^{*}$&$\dot{\rm J}^{*}$&log J$_{rel}^{*}$\\ 
			&(M$_\odot$)&(M$_\odot$)&(R$_\odot$)&(R$_\odot$)&(L$_\odot$)&(L$_\odot$)&($\times10^{51}$)&($\times10^{42}$)&\\
			\hline
			V48&1.07$\pm$0.02&0.16$\pm$0.01 &1.09$\pm$0.02 &0.49$\pm$0.01 &1.36$\pm$0.03 &0.27$\pm$0.01 &1.32$\pm$0.04&-2.22$\pm$0.02&-0.97$\pm$0.02\\            V49&0.82$\pm$0.03&0.97$\pm$0.03&0.84$\pm$0.03&0.90$\pm$0.04&0.55$\pm$0.02&0.64$\pm$0.03&5.46$\pm$0.02&-1.30$\pm$0.02&-0.36$\pm$0.01\\			V51&0.79$\pm$0.04&0.99$\pm$0.05&0.84$\pm$0.04&0.93$\pm$0.04&0.57$\pm$0.03&0.69$\pm$0.04&5.42$\pm$0.03&-1.37$\pm$0.01&-0.36$\pm$0.01\\			V55&0.81$\pm$0.02&1.01$\pm$0.03&0.86$\pm$0.02&0.94$\pm$0.02&0.61$\pm$0.02&0.73$\pm$0.02&5.63$\pm$0.02&-1.42$\pm$0.03&-0.34$\pm$0.03\\			NV4&0.89$\pm$0.02&1.09$\pm$0.03&0.95$\pm$0.02&1.04$\pm$0.03&0.80$\pm$0.02&0.94$\pm$0.03&6.72$\pm$0.09&-1.66$\pm$0.04&-0.27$\pm$0.03\\
			\hline
		\end{tabular}
		\label{table4}
		\begin{tablenotes}
			\footnotesize
			\item {$^*$}\textbf{in cgs units.} 
		\end{tablenotes}
	\end{table}
	\section{Results}
	
	\subsection{NV4}
	On the basis of light curve plotted from the observations, NV4 was identified as a W Ursae Majoris-type eclipsing variable, discovered among 19 new variables in Wise Observatory study \citep{safonova2016search}. From the detailed photometric analysis, performed on the collected data, we find that NV4 is a high-mass ratio (q$\sim$1.2) marginal (f$\sim$20\%) contact binary . There isn't any period analysis reported for NV4 and there isn't any previous data available in the literature as well, since it is being reported as a contact binary for the first time. 
	
	Photometric observations exhibit a difference between the two maxima in the light curve, also, called as the O’Connell effect (\cite{o1951so}, \cite{davidge1984study}). The best fit light curve was obtained only after incorporating a hot spot on the secondary component. Chromospheric activity is the most common justification given for these observed light curve asymmetries. This activity can be due to the underlying magnetic activity of the system manifested by surface phenomena like star spots, flares, stellar winds, etc. Hot spots on the surface of stars are generally attributed to the flares or impact of mass-transferring gas streams (\cite{{wilsey2009revisiting},{kouzuma2019starspots}}). From the estimated absolute parameters, to understand the equilibrium state of the system, the relative orbital angular momentum of the system is determined and found to be $\log J_{rel}$$\sim$-0.25 which is higher than the values defined for contact binaries ($\log J_{rel}$$\sim$-0.5) \citep{popper1977evolutionary}. This generally indicates that the system is in a stable configuration. The orbital angular momentum (AM) of the system is about $7\times10^{51}$ cgs units and the AML from the system is $1.64 \times10^{42}$ cgs units. On the basis of above values, we propose that the system is undergoing the Thermal Relaxation Oscillation (TRO) Cycle. To understand the evolutionary state of the system, Log M vs Log R (M-R) diagram is plotted in Fig.~\ref{figure4}. The Zero Age Main Sequence (ZAMS) and Terminal Age Main Sequence (TAMS) lines of Fig.~\ref{figure4} were adopted from \cite{ma2022photometric}. We observe that both components of the system are above the ZAMS line and are under-luminous for their main-sequence masses. Hence, continuous and high-precision photometric data is highly needed to conduct the period variation analysis study, to interpret the present and future evolutionary scenario of the binary more accurately.
	
	\subsection{V48}
	The variable V48 was catalogued as a contact binary and an X-ray source by \cite{kaluzny2013cluster} and \cite{bassa2004x}. It appears as a blend of three or four stars, signifying that the X-ray source coincides with the optical variables. The light curve and solutions obtained show that it is a totally eclipsing (i$\sim$$80^{o}$), low mass ratio (q$\sim$0.15) contact binary with deep contact nature ($f$$\sim$51\%). The system exhibits a long-term period decrease with $\dot P=-6.56 \times 10^{-8}$ d/yr, which is generally attributed to mass transfer from a more massive to a less massive component or by AML or both. To know the reason for the long-term period decrease in V48, the mass transfer rate was estimated using the following equation,
	\begin{equation}
	\frac{\dot P}{P}= -3\dot M_1\left[{\frac{1}{M_1}}-{\frac{1}{M_2}}\right]\,.
	\end{equation}
	The mass transfer rate of $\dot M_1= -3.18\times10^{-8}$ M$_{\odot}$/yr indicates that the more massive primary is losing the mass and it can be one of the reasons for the orbital period decrease. However, we cannot rule out the possibility of AML via stellar wind as this star was classified as a magnetically active binary (\cite{{kaluzny2013cluster},{bassa2004x}}) as evident by the presence of hot spots in the derived light curve solutions in the current work. From Fig.~\ref{figure4}, we observe that the primary is above ZAMS and the secondary is above TAMS, indicating that the secondary has evolved faster and has left TAMS.
	
	\subsection{V49}
	V49 also has been classified as a magnetically active X-ray source (\cite{kaluzny2013cluster}; \cite{bassa2004x}) and from the light curve solution obtained it is an over-contact binary with a fill-out factor of $\sim50$\% and a high mass ratio of 1.2. The evolutionary theories of contact binaries suggest that these systems with high mass ratios would have recently evolved into contact \cite{hrivnak1989high} and are in the AML-controlled stage (\cite{{qian2001possible},{qian2003overcontact}}). It is experiencing a secular decrease in orbital period at a rate of $\dot P =-3.75\times10^{-8}$ d/yr, which may be due to mass transfer from more massive component primary to less massive secondary. We can confirm this from the estimated mass transfer rate $\dot M_1=-1.00\times10^{-7} M_{\odot}$/yr. The negative sign indicates that the more massive primary is losing mass. The system may evolve from contact to over-contact configuration and be a progenitor for a single fast-rotating star. Further, the period variation showed a cyclic variation with a semi-amplitude $\sim0.0005$ day and period about 9 years. Such cyclic variation can be attributed to the presence of a third body which cause the light time travel effect (LTTE) (\cite{borkovits1996invisible}) or to the magnetic activity cycle, i.e., Applegate Mechanism (\cite{applegate1992mechanism}). The possible reason for the observed cyclic variation is briefly discussed under Section \ref{sec:discussion}. The estimated absolute masses plotted in Fig.~\ref{figure4} show the positions of primary and secondary are slightly above ZAMS, respectively indicating that the primary and secondary components are moderately evolved.
	\subsection{V51}
	V51 has been classified as a contact binary by \cite{kaluzny1997ccd}. The photometric solutions indicate that it is a shallow contact binary with a fill-out factor of $\sim$30\%. The mass ratio is relatively high around 1.46 and the period change analysis indicates it is undergoing secular decrease at a rate of $\dot P =-4.05\times 10^{-7}$ d/yr, which can be due to mass transfer or AML. Assuming the mass transfer is conservative, its  rate was calculated to be $\dot M_2 =-1.68\times10^{-6} M_{\odot}$/yr suggesting the massive secondary is losing its mass. The positions of the primary and secondary components is above the ZAMS line in Fig.~\ref{figure4} indicating moderately evolved nature of the components.
	\subsection{V55}
	V55 also is one of the contact binaries categorized as a magnetically active X-ray source (\cite{kaluzny2013cluster}; \cite{bassa2004x}). It is a high mass ratio contact binary with $q=1.34$ with shallow contact configuration $\sim$24\%, similar to NV4. The period increase observed in the (O-C) diagram at a $\dot P =4.47\times10^{-9}$ d/yr, can be attributed to the mass transfer between primary and secondary components. The mass transfer rate calculated is $\dot M_1=-3.36\times10^{-7}$ M$_{\odot}$/yr, indicating that the primary is losing mass to the secondary. This rate seems to be quite high for contact binaries however, it can be verified through further observational investigations. The position of the components of V55 in Fig.~\ref{figure4} shows that they are just above the ZAMS indicating that they are slowly evolving.
	\begin{figure*}
		\centering
		\centering
		\includegraphics[width=0.75\textwidth]{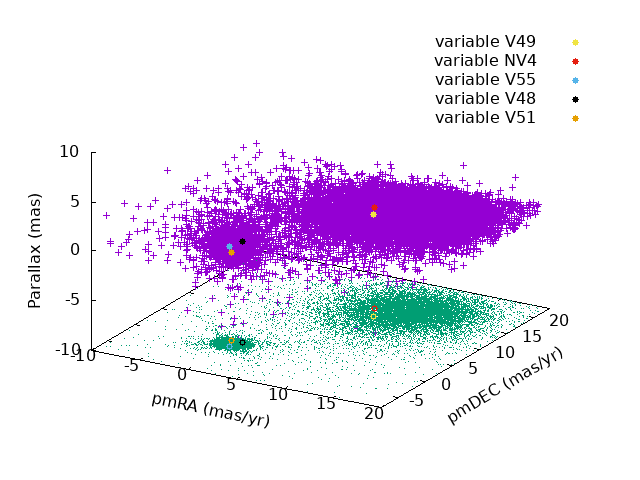}
		\caption{Analysis of M4 cluster membership of 33433 stars with measured relative parallaxes (in milliarcseconds) and proper motions (in milliarcseconds per year) located within the tidal radius of the cluster: 3-parameter VPD.}
		\label{figure6}
	\end{figure*}
	\begin{figure*}
		\centering
		\includegraphics[width=0.48\textwidth]{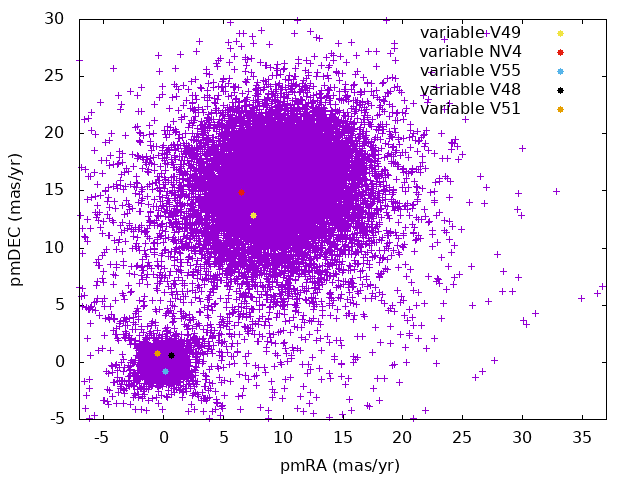}
		\includegraphics[width=0.48\textwidth]{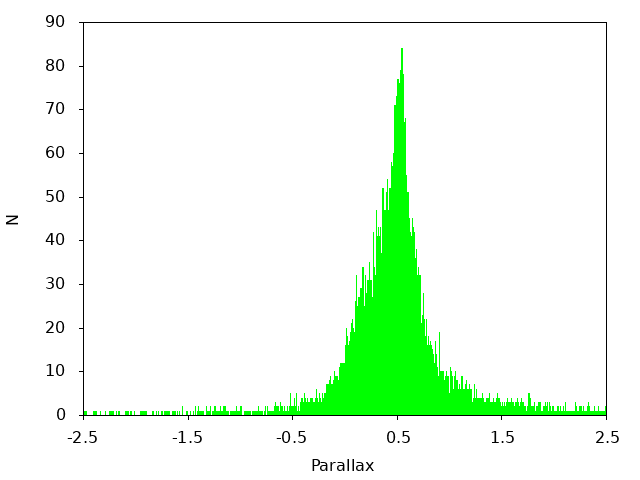}
		\caption{{\it Top}: 2-dimensional VPD. {\it Bottom}: the histogram plot of parallaxes for these stars, with the mean value of 0.5517.}
		\label{figure7}
	\end{figure*}
	\begin{figure*}
		\centering
		\includegraphics[width=0.9\textwidth]{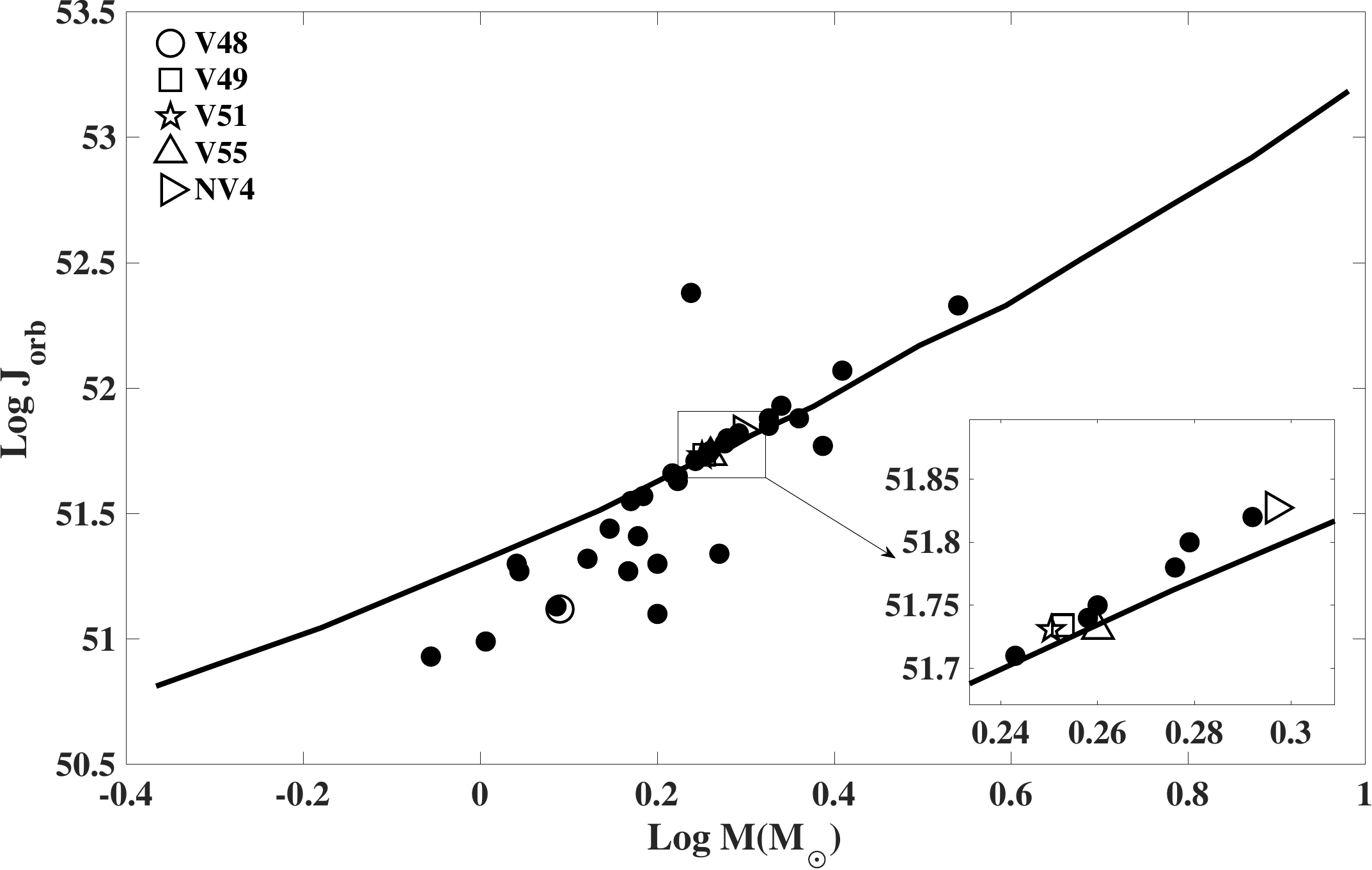}
		\caption{The relationship between Log M vs. Log J$_{orb}$ of various contact binaries.}
		\label{figure8}
	\end{figure*}
	\begin{figure*}
		\centering
			\includegraphics[width=0.95\textwidth]{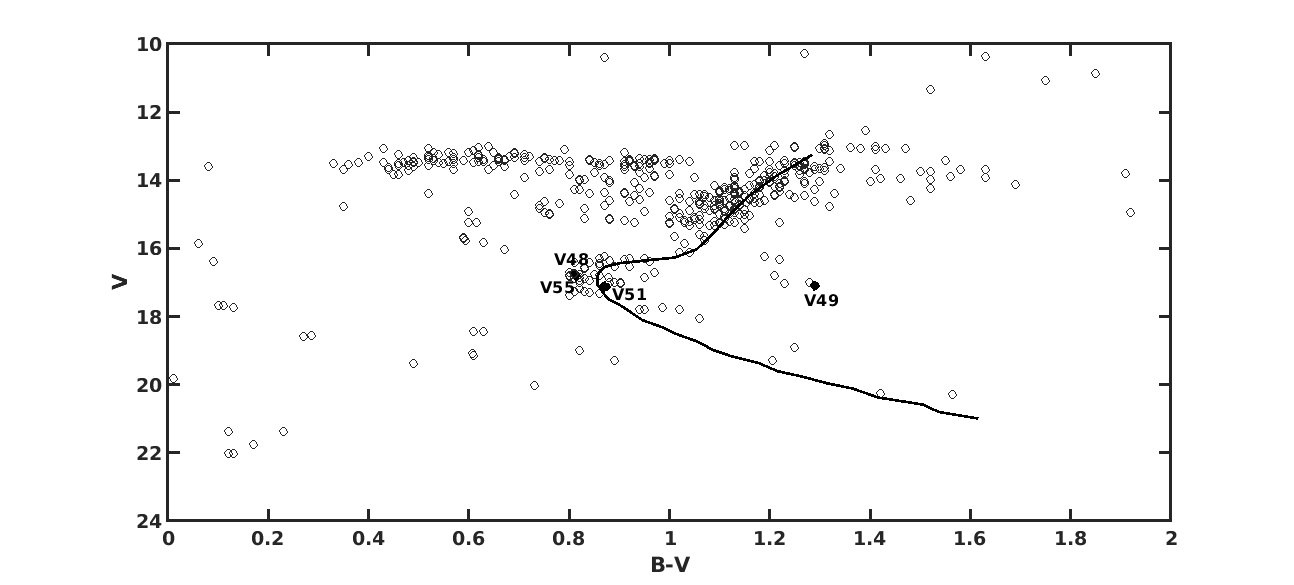}
			\includegraphics[width=0.8\textwidth]{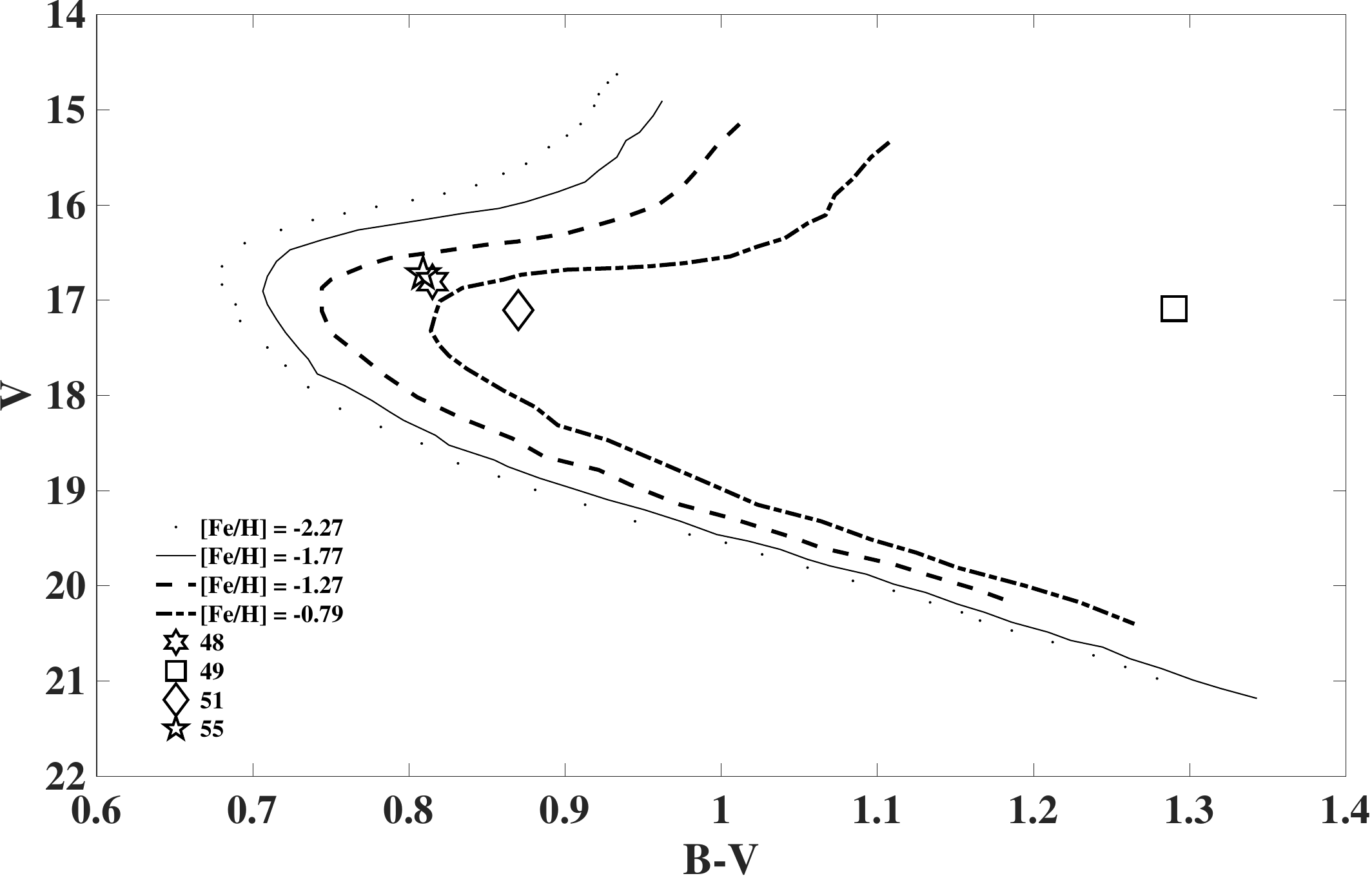}
		\caption{B-V vs. V of the M4 globular cluster. The top panel displays a different catalog of variables, whereas the bottom panel displays the four variables along with a model of a solid line at various metallicities. The solid lines were obtained by applying 16 Gyr isochrones for -2.37<[Fe/H]<-0.79 by \cite{kanatas1995ccd}.}
		\label{figure9}
	\end{figure*}
	\section{Discussion}
	\label{sec:discussion}
	
	Out of the four binaries in the study for which B and V bands data are available, V48 is a totally eclipsing, over-contact binary with a low mass ratio (q $\sim$ 0.15) and high fillout factor (f$\sim$50\%), and such objects are the progenitors for FK Com-type and blue straggler stars. These systems show gradual decrease in mass ratio and in the period leading to shrinking of critical Roche lobes, thus increasing fill-out factor evolving into single fast-rotating stars. This is verified using Fig.~\ref{figure5} which shows the period–colour relation. The solid line in Fig.~\ref{figure5} shows the short-period blue envelope (SPBE) theoretical line, obtained using the relation given by \cite{rucinski1997absolute},
	\begin{equation}
	(B-V)_{SPBE}=0.04\times P^{-2.25}\,.
	\end{equation}
	It can be observed that V48 lies above the SPBE line and the rest all lie below the SPBE line indicating that it is an interesting source that is moving towards longer periods and cooler temperatures and a potential member to undergo a merger. Further, from Fig.~\ref{figure4}, we observe that the less massive components of contact binaries in GCs are more evolved than their more massive companions, indicating that less massive components are under-sized compared to their companions that are mostly in the main sequence. To investigate further the contact state of all the binaries in the study, the orbital AM was calculated, using the equation given by \cite{christopoulou2013extensive}, and compared with the sample of contact binaries that are well-studied in various GCs (displayed in Table \ref{table5}).
	\begin{equation}
	J_{orb} = 1.24\times10^{52}\times M^{(5/3)}\times P^{(1/3)}\times q\times(1+q)^{-2}\,,
	\end{equation}
	where $M$ is the total mass, $P$ is the period and $q$ is the mass ratio obtained from the photometry of the systems. Fig.~\ref{figure8} shows the relationship between the $\log J_{orb}$ and $\log M(M_{\odot}$), where the solid line represents a border that separates the contact zone from that of the detached, obtained by \cite{eker2006dynamical}. We observed that the variable V48 is in the contact region and other variables (V49, V51, V55, and NV4) have just entered the contact region. The variable V49 could be oscillating between contact and detached phases exhibiting Thermal Relaxation Oscillation (TRO) cycles (as in Case 4 of Table~5 in \cite{pothuneni2023first}). 
	Among the four binaries with period variation, V49 exhibited a cyclic variation in addition to the period decrease. The period change may be attributed to the presence of a third body or a magnetic activity cycle or both (\cite{pothuneni2023first}). In order to verify the cause, the quadruple moment (QM) of both the components of V49 is determined using the relation given by \cite{lanza2002gravitational}
	\begin{equation}
	\Delta Q = -\frac{\Delta P}{P}\times\frac{Ma^2}{9}
	\end{equation}
	where $\Delta$P is observed orbital period variation, M is the mass of the active star and \begin{math} a
	\end{math} is the semi-major axis of the binary orbit. The calculated QMs for both the components of V49 were found to be much less than that of typical values for active stars ($10^{51-53}$g cm$^2$). Hence, the QM is not enough to produce the observed cyclic change and presence of a third body would be a feasible reason. From the sine fit, the third body parameters were derived using relations given by \cite{borkovits1996invisible} and are found to be: P\begin{math}_{3}\sim9\end{math} years, m\begin{math}_{3(min)}\sim31\end{math} M$_{Jupiter}$ and a\begin{math}_{3}\sim5\end{math} AU. The third body could be a brown dwarf. Though third light is not found in the LC analysis of V49 which can be due to a faint third body. However, continuous long-term high-precision photometric observations in the future may help in affirming the same.\\
	Fig.~\ref{figure9} (top panel) represents the CMD plotted using the available data of all variable stars studied in the M4 Cluster. The data was collected from \cite{kaluzny2013cluster}, \cite{stetson2014optical}, \cite{mochejska2002clusters}, \cite{alcaino1988bvi}, \cite{zloczewski2013proper}, \cite{alcaino1997main} and \cite{cudworth1990astrometry}. The magnitude of variables collected lies between 10<V<22. The binaries in the study are overplotted and it is observed that V48, V51, and V55 are placed well on the main sequence, and V49 is observed to be evolving towards it. To understand the age and metallicity of the variables in the study, the best-matched theoretical isochrone and metallicity are fitted over a magnitude-color relation diagram (the bottom panel of Fig.~\ref{figure9}) adopted from \cite{kanatas1995ccd}. It clearly shows that V48, V51, and V55 are at the beginning of the plateau region and are in line with the predicted cluster age ($16\pm1$ Gyr) and metallicity $\left[\frac{Fe}{H}\right]=-1.27$ as given by \cite{kanatas1995ccd}. 
	
	Variable V49 is an interesting object with an inclination $\sim$86$^{\circ}$ and totality in the light curve at the primary minima, and is a typical variable when compared to the other three cluster sources, but due to its discrepant value of the distance modulus \cite{kaluzny1997ccd} and outlier position on the CMD (Fig. \ref{figure9}), there is a question of whether it is a member of the M4 cluster. It is a well-known fact that M4 has a proper motion that is well separated from the field stars \citep{Wallace2018}, which makes it a useful determinant of cluster membership. We have used the recent measurements from Gaia Data Release 3 (Gaia DR3; \citet{GAIA2022}) to verify cluster membership of the variables under study. Gaia DR3 provides high-precision proper motion data for more than a billion sources in terms of several astrometric parameters: coordinates ($\a,\d$), parallaxes ($\pi$), and proper motion in right ascension and declination ($\mu_{\a}\cos{\d},\mu_{\d}$). 
	The source data were downloaded from Gaia DR3 database service\footnote{https://vizier.cds.unistra.fr/viz-bin/VizieR-3?-source=I/355/gaiadr3} for all sources lying within the tidal radius of the cluster ($32.49’$; \citet{Baumgardt2019}), brighter than $G$=19 \citep{Wallace2018} and having parallax $<$5. This resulted in 33433 stars with measured both proper motions and parallaxes. To enhance the cluster data, we present the data in three-dimensional parameter space ($\pi,\mu_{\a}\cos{\d},\mu_{\d}$) in Figures \ref{figure6} and \ref{figure7} (Top panel), where the zero point is the mean proper motion and mean parallax of the cluster ($-12.48$, $-18.99$, and 0.556, respectively; \cite{Baumgardt2019}). Here, the bulk of stars clustered around the origin of the 3D VPD consists mostly of cluster members, while field stars form well-defined clump that is well-separated from the cluster stars. The data for variables under the study (i.e., NV4, V48, V49, V51 and V55) was also included in the same dataset; they are indicated in Figures \ref{figure6} and \ref{figure7} by points with different colours. It is seen that variables V49 and NV4 are not cluster members; though NV4 has the same relative parallax as other variables, its proper motion places it in the field stars clump. Parallax distribution of all downloaded stars is shown in the bottom panel of Fig. \ref{figure7}.
	\section{Conclusions}
	
	A photometric study was carried out on five contact binaries, and the period variation analysis was performed on four contact binaries in the globular cluster M4. From the sample of well-studied close binaries in other globular clusters, these systems typically appear in the period interval 0.27<P<0.9 days and are very few at longer periods and bluer colours. A secular period change is observed to be associated among the four binaries in the current study which is often attributed to the mass transfer or to the angular momentum loss due to magnetic stellar wind in these systems. V48 is a deep, low-mass ratio over-contact binary which with its mass, angular momentum loss, and orbital period decrease may finally merge into a rapid-rotating single star. Because of the  observed totality in the light curve, we find the photometric solutions obtained reliable (\cite{li2021photometric}, \cite{mochnacki1972model}, \cite{devarapalli2020comprehensive}, \cite{jagirdar2023first}). V49 is a deep, high-mass ratio over-contact binary, in which LC exhibits totality at the primary minima, and LC solutions converge towards an inclination of $\sim$$86^{o}$. The other variables in the study are in the marginal contact phase and are partially eclipsing binaries. V48, V51, and V55 are cluster members, while NV4 and V49 are the field stars. All the components of the binaries in the study are observed to follow the main sequence trend and are located within ZAMS and TAMS limits. Long-term monitoring of a variety of such short-period contact binary systems in clusters, with respect to their orbital parameters and period variation studies, may lead to a better understanding of the evolution of such systems in clusters.
	\begin{table*}
			\centering
			\caption{Catalog of W UMa type binaries from various globular clusters.}
			\label{table5}
			\scriptsize
			\begin{tabular}{p{1.2cm}p{3.5cm}p{0.825cm}p{0.825cm}p{0.80cm}p{0.9cm}p{0.80cm}p{0.80cm}p{0.80cm}p{2.0cm}}
				\hline
				Cluster&Variable&P&logM$_1$&logM$_2$&logR$_1$&logR$_2$&logJ$_{orb}$&References\\
				&&(d)&(M$_{\odot}$)&(M$_{\odot}$)&(R$_{\odot}$)&(R$_{\odot}$)&(cgs units)&\\
				\hline
				47 Tuc &V95&0.278899&0.02&-0.77&0.03&-0.31&51.13&1\\
				\hline
				M13 &V57&0.285416&-0.02&-0.16&-0.05&-0.11&51.66&2\\
				\hline
				M15 &W1&0.233060&-0.14&-0.43&-0.03&-0.15&51.30&3\\
				&W2&0.235760&0.08&0.04&0.03&0.03&51.88&3\\
				\hline
				M35 &2MASS J06092044 +2415155&0.529600&0.18&0.03&0.20&0.13&52.07&4\\
				\hline
				M4 &V47&0.269973&0.22&-0.70&0.09&-0.30&51.34&5\\
				&V53&0.308512&0.17&-0.92&0.14&-0.32&51.10&6\\
				&V48&0.282686&0.03&-0.79&0.04&-0.31&51.11&7\\
				&V49&0.297444&-0.08&-0.01&-0.08&-0.05&51.73&7\\
				&V51&0.303683&-0.10&0.004&-0.08&-0.03&51.73&7\\
				&V55&0.310703&-0.09&0.003&-0.07&-0.02&51.75&7\\
				&NV4&0.344440&-0.05&0.04&-0.02&0.02&51.83&7\\
				\hline
				M54 &V134&0.909521&0.36&0.08&0.43&0.32&52.33&8\\
				&V144&0.721590&0.32&-0.47&0.41&0.08&51.77&8\\
				\hline
				M67 &AH Cnc&0.360441&0.11&-0.73&0.14&-0.22&51.27&4\\
				\hline
				M71 &2MASS J19532554 + 1851175&0.357907&0.09&-0.57&0.11&-0.17&51.41&4\\
				&2MASS J19533427 + 1844047&0.404341&0.15&-0.74&0.19&-0.19&51.30&4\\
				\hline
				NGC 2539 &V3&0.292000&-0.01&-0.11&-0.05&-0.09&51.71&9\\
				&V4&0.339999&0.03&0.03&0.002&0.002&51.88&9\\
				\hline
				NGC 2632 &V1&0.291000&-0.03&0.08&-0.24&-0.29&51.85&10\\
				\hline
				NGC 5139&V206&0.307351&-0.05&0.003&-0.06&-0.04&51.78&11\\
				&V207&0.275867&-0.28&-0.02&-0.16&-0.05&51.55&11\\
				&V208&0.305526&0.02&-0.31&0.003&-0.14&51.57&12\\
				&V214&0.341804&-0.07&0.02&-0.03&0.02&51.80&12\\
				&V236&0.296318&-0.01&-0.07&-0.05&-0.07&51.74&12\\
				&V240&0.331884&-0.04&0.02&-0.03&0.002&51.82&12\\
				\hline
				NGC 5466&UW CVn&0.292457&-0.59&0.03&-0.24&0.03&51.32&13\\
				&NH 19&0.342144&0.05&-0.26&0.05&-0.09&51.65&14\\
				&NH 30&0.297536&0.02&-0.44&0.01&-0.19&51.44&14\\
				\hline
				NGC 6341&V798 Her&0.295111&-0.02&-0.07&-0.05&-0.07&51.75&15\\
				\hline
				NGC 6397&V7&0.269861&-0.52&-0.09&-0.27&-0.07&51.27&16\\
				&V8&0.271243&-0.06&-0.86&0.004&-0.34&50.99&16\\
				\hline
				NGC 6791 &V1&0.267700&-0.12&-0.05&-0.12&-0.09&51.66&17\\
				\hline
				NGC 6866 &ID487&0.415110&0.09&1.98&0.09&0.05&53.38&18\\
				&ID494&0.366709&0.08&-0.33&0.09&-0.09&51.63&18\\
				\hline
			\end{tabular}\\[0.1in]
			\begin{tablenotes}
				\footnotesize
				\item 1.\cite{liu2014identification}; 2. \cite{deras2019new}, 3. \cite{saad2005light}, 4. \cite{koccak2020photometric}, 5. \cite{liu2011two}, 6. \cite{li2017comp}, 7. Current Study, 8. \cite{li2013one}, 9. \cite{kiron2012photometric}, 10. \cite{priya2013photometric}, 11. \cite{li2018photometric}, 12. \cite{kiron2011photometric}, 13. \cite{kopacki1995uw}, 14. \cite{kallrath1992modeling}, 15.\cite{anees2021differential}, 16.\cite{li2013two}, 17.\cite{rukmini2005photometric}, 18.\cite{joshi2016photometric}.
		\end{tablenotes}
	\end{table*}
	
	\printcredits
	
	\section*{Acknowledgements}
	
	MS acknowledges the financial support by the DST, Government of India, under the Women Scientist Scheme (PH) project reference number SR/WOS-A/PM-17/2019.

	\bibliographystyle{cas-model2-names}
	
	\bibliography{cas-refs.bib}

\begin{thebibliography}{70}
\expandafter\ifx\csname natexlab\endcsname\relax\def\natexlab#1{#1}\fi
\providecommand{\url}[1]{\texttt{#1}}
\providecommand{\href}[2]{#2}
\providecommand{\path}[1]{#1}
\providecommand{\DOIprefix}{doi:}
\providecommand{\ArXivprefix}{arXiv:}
\providecommand{\URLprefix}{URL: }
\providecommand{\Pubmedprefix}{pmid:}
\providecommand{\doi}[1]{\href{http://dx.doi.org/#1}{\path{#1}}}
\providecommand{\Pubmed}[1]{\href{pmid:#1}{\path{#1}}}
\providecommand{\bibinfo}[2]{#2}
\ifx\xfnm\relax \def\xfnm[#1]{\unskip,\space#1}\fi
\bibitem[{Alcaino et~al.(1988)Alcaino, Liller and Alvarado}]{alcaino1988bvi}
\bibinfo{author}{Alcaino, G.}, \bibinfo{author}{Liller, W.},
  \bibinfo{author}{Alvarado, F.}, \bibinfo{year}{1988}.
\newblock \bibinfo{title}{Bvi ccd photometry of the globular cluster m4}.
\newblock \bibinfo{journal}{ApJ} \bibinfo{volume}{330}, \bibinfo{pages}{569}.
\bibitem[{Alcaino et~al.(1997)Alcaino, Liller, Alvarado, Kravtsov, Ipatov,
  Samus and Smirnov}]{alcaino1997main}
\bibinfo{author}{Alcaino, G.}, \bibinfo{author}{Liller, W.},
  \bibinfo{author}{Alvarado, F.}, \bibinfo{author}{Kravtsov, V.},
  \bibinfo{author}{Ipatov, A.}, \bibinfo{author}{Samus, N.},
  \bibinfo{author}{Smirnov, O.}, \bibinfo{year}{1997}.
\newblock \bibinfo{title}{The main sequence of the globular cluster m4= ngc
  6121 from ccd ntt photometry}.
\newblock \bibinfo{journal}{AJ} \bibinfo{volume}{114}, \bibinfo{pages}{189}.
\bibitem[{Anees et~al.(2021)Anees, Goderya and Khan}]{anees2021differential}
\bibinfo{author}{Anees, K.}, \bibinfo{author}{Goderya, S.N.},
  \bibinfo{author}{Khan, F.M.}, \bibinfo{year}{2021}.
\newblock \bibinfo{title}{Differential photometry of eclipsing binary system
  v798 her in globular cluster ngc 6341}.
\newblock \bibinfo{journal}{JAAVSO} \bibinfo{volume}{49}, \bibinfo{pages}{116}.
\bibitem[{Applegate(1992)}]{applegate1992mechanism}
\bibinfo{author}{Applegate, J.H.}, \bibinfo{year}{1992}.
\newblock \bibinfo{title}{A mechanism for orbital period modulation in close
  binaries}.
\newblock \bibinfo{journal}{Astrophysical Journal, Part 1 (ISSN 0004-637X),
  vol. 385, Feb. 1, 1992, p. 621-629.} \bibinfo{volume}{385},
  \bibinfo{pages}{621--629}.
\bibitem[{Bassa et~al.(2004)Bassa, Pooley, Homer, Verbunt, Gaensler, Lewin,
  Anderson, Margon, Kaspi and van~der Klis}]{bassa2004x}
\bibinfo{author}{Bassa, C.}, \bibinfo{author}{Pooley, D.},
  \bibinfo{author}{Homer, L.}, \bibinfo{author}{Verbunt, F.},
  \bibinfo{author}{Gaensler, B.M.}, \bibinfo{author}{Lewin, W.H.},
  \bibinfo{author}{Anderson, S.F.}, \bibinfo{author}{Margon, B.},
  \bibinfo{author}{Kaspi, V.M.}, \bibinfo{author}{van~der Klis, M.},
  \bibinfo{year}{2004}.
\newblock \bibinfo{title}{X-ray sources and their optical counterparts in the
  globular cluster m4}.
\newblock \bibinfo{journal}{ApJ} \bibinfo{volume}{609}, \bibinfo{pages}{755}.
\bibitem[{{Baumgardt} et~al.(2019){Baumgardt}, {Hilker}, {Sollima} and
  {Bellini}}]{Baumgardt2019}
\bibinfo{author}{{Baumgardt}, H.}, \bibinfo{author}{{Hilker}, M.},
  \bibinfo{author}{{Sollima}, A.}, \bibinfo{author}{{Bellini}, A.},
  \bibinfo{year}{2019}.
\newblock \bibinfo{title}{Mean proper motions, space orbits, and velocity
  dispersion profiles of galactic globular clusters derived from gaia dr2
  data}.
\newblock \bibinfo{journal}{MNRAS} \bibinfo{volume}{482},
  \bibinfo{pages}{5138--5155}.
\newblock \DOIprefix\doi{10.1093/mnras/sty2997},
  \href{http://arxiv.org/abs/1811.01507}{\tt arXiv:1811.01507}.
\bibitem[{Borkovits and Heged{\"u}s(1996)}]{borkovits1996invisible}
\bibinfo{author}{Borkovits, T.}, \bibinfo{author}{Heged{\"u}s, T.},
  \bibinfo{year}{1996}.
\newblock \bibinfo{title}{On the invisible components of some eclipsing
  binaries}.
\newblock \bibinfo{journal}{Astronomy and Astrophysics Supplement Series}
  \bibinfo{volume}{120}, \bibinfo{pages}{63--75}.
\bibitem[{Christopoulou and Papageorgiou(2013)}]{christopoulou2013extensive}
\bibinfo{author}{Christopoulou, P.E.}, \bibinfo{author}{Papageorgiou, A.},
  \bibinfo{year}{2013}.
\newblock \bibinfo{title}{An extensive analysis of the triple w uma type binary
  fi boo}.
\newblock \bibinfo{journal}{AJ} \bibinfo{volume}{146}, \bibinfo{pages}{157}.
\bibitem[{Cudworth and Rees(1990)}]{cudworth1990astrometry}
\bibinfo{author}{Cudworth, K.}, \bibinfo{author}{Rees, R.},
  \bibinfo{year}{1990}.
\newblock \bibinfo{title}{Astrometry and photometry in the globular cluster
  m4}.
\newblock \bibinfo{journal}{AJl} \bibinfo{volume}{99}, \bibinfo{pages}{1491}.
\bibitem[{Davidge and Milone(1984)}]{davidge1984study}
\bibinfo{author}{Davidge, T.}, \bibinfo{author}{Milone, E.},
  \bibinfo{year}{1984}.
\newblock \bibinfo{title}{A study of the o'connell effect in the light curves
  of eclipsing binaries}.
\newblock \bibinfo{journal}{Astrophysical Journal Supplement Series (ISSN
  0067-0049), vol. 55, Aug. 1984, p. 571-584. Research supported by the
  University Grants Committee, Natural Sciences and Engineering Research
  Council, and University of Calgary.} \bibinfo{volume}{55},
  \bibinfo{pages}{571--584}.
\bibitem[{Deras et~al.(2019)Deras, Arellano~Ferro, L{\'a}zaro, Bustos~Fierro,
  Calder{\'o}n, Muneer and Giridhar}]{deras2019new}
\bibinfo{author}{Deras, D.}, \bibinfo{author}{Arellano~Ferro, A.},
  \bibinfo{author}{L{\'a}zaro, C.}, \bibinfo{author}{Bustos~Fierro, I.},
  \bibinfo{author}{Calder{\'o}n, J.}, \bibinfo{author}{Muneer, S.},
  \bibinfo{author}{Giridhar, S.}, \bibinfo{year}{2019}.
\newblock \bibinfo{title}{A new study of the variable star population in the
  hercules globular cluster (m13; ngc 6205)}.
\newblock \bibinfo{journal}{MNRAS} \bibinfo{volume}{486},
  \bibinfo{pages}{2791--2808}.
\bibitem[{Devarapalli et~al.(2020)Devarapalli, Jagirdar, Prasad, Thomas, Ahmed,
  Gralapally and Das}]{devarapalli2020comprehensive}
\bibinfo{author}{Devarapalli, S.}, \bibinfo{author}{Jagirdar, R.},
  \bibinfo{author}{Prasad, R.}, \bibinfo{author}{Thomas, V.},
  \bibinfo{author}{Ahmed, S.}, \bibinfo{author}{Gralapally, R.},
  \bibinfo{author}{Das, J.}, \bibinfo{year}{2020}.
\newblock \bibinfo{title}{Comprehensive study of a neglected contact binary tyc
  5532-1333-1}.
\newblock \bibinfo{journal}{Monthly Notices of the Royal Astronomical Society}
  \bibinfo{volume}{493}, \bibinfo{pages}{1565--1573}.
\bibitem[{Eker et~al.(2006)Eker, Demircan, Bilir and
  Karata{\c{s}}}]{eker2006dynamical}
\bibinfo{author}{Eker, Z.}, \bibinfo{author}{Demircan, O.},
  \bibinfo{author}{Bilir, S.}, \bibinfo{author}{Karata{\c{s}}, Y.},
  \bibinfo{year}{2006}.
\newblock \bibinfo{title}{Dynamical evolution of active detached binaries on
  the log j o--log m diagram and contact binary formation}.
\newblock \bibinfo{journal}{MNRA} , \bibinfo{pages}{1483}.
\bibitem[{{Gaia Collaboration}(2022)}]{GAIA2022}
\bibinfo{author}{{Gaia Collaboration}}, \bibinfo{year}{2022}.
\newblock \bibinfo{title}{Vizier online data catalog: Gaia dr3 part 1. main
  source (gaia collaboration, 2022)}.
\newblock \bibinfo{journal}{VizieR Online Data Catalog} ,
  \bibinfo{pages}{I/355}.
\bibitem[{Gazeas(2009)}]{gazeas2009physical}
\bibinfo{author}{Gazeas, K.}, \bibinfo{year}{2009}.
\newblock \bibinfo{title}{Physical parameters of contact binaries through 2-d
  and 3-d correlation diagrams}.
\newblock \bibinfo{journal}{Comm. in Asteroseismology} \bibinfo{volume}{159},
  \bibinfo{pages}{129}.
\bibitem[{Hrivnak(1989)}]{hrivnak1989high}
\bibinfo{author}{Hrivnak, B.J.}, \bibinfo{year}{1989}.
\newblock \bibinfo{title}{The unusual, high-mass-ratio contact binary vz
  piscium}, in: \bibinfo{booktitle}{International Astronomical Union
  Colloquium}, \bibinfo{organization}{Cambridge University Press}. p.
  \bibinfo{pages}{348}.
\bibitem[{Hut et~al.(1992)Hut, McMillan, Goodman, Mateo, Phinney, Pryor,
  Richer, Verbunt and Weinberg}]{hut1992binaries}
\bibinfo{author}{Hut, P.}, \bibinfo{author}{McMillan, S.},
  \bibinfo{author}{Goodman, J.}, \bibinfo{author}{Mateo, M.},
  \bibinfo{author}{Phinney, E.}, \bibinfo{author}{Pryor, C.},
  \bibinfo{author}{Richer, H.B.}, \bibinfo{author}{Verbunt, F.},
  \bibinfo{author}{Weinberg, M.}, \bibinfo{year}{1992}.
\newblock \bibinfo{title}{Binaries in globular clusters}.
\newblock \bibinfo{journal}{PASP} \bibinfo{volume}{104}, \bibinfo{pages}{981}.
\bibitem[{Jagirdar et~al.(2023)Jagirdar, Pothuneni, Devarapalli, Manurkar
  et~al.}]{jagirdar2023first}
\bibinfo{author}{Jagirdar, R.}, \bibinfo{author}{Pothuneni, R.R.},
  \bibinfo{author}{Devarapalli, S.P.}, \bibinfo{author}{Manurkar, B.}, et~al.,
  \bibinfo{year}{2023}.
\newblock \bibinfo{title}{The first photometric, period variation, and flare
  activity studies of tic 24233294: Reviewing with the latest results of
  well-studied late-type active binaries}.
\newblock \bibinfo{journal}{Advances in Space Research} .
\bibitem[{Joshi et~al.(2016)Joshi, Jagirdar and Joshi}]{joshi2016photometric}
\bibinfo{author}{Joshi, Y.C.}, \bibinfo{author}{Jagirdar, R.},
  \bibinfo{author}{Joshi, S.}, \bibinfo{year}{2016}.
\newblock \bibinfo{title}{Photometric studies of two w uma type variables in
  the field of distant open cluster ngc 6866}.
\newblock \bibinfo{journal}{Research in Astronomy and Astrophysics}
  \bibinfo{volume}{16}, \bibinfo{pages}{011}.
\bibitem[{Kallrath et~al.(1992)Kallrath, Milone and
  Stagg}]{kallrath1992modeling}
\bibinfo{author}{Kallrath, J.}, \bibinfo{author}{Milone, E.},
  \bibinfo{author}{Stagg, C.}, \bibinfo{year}{1992}.
\newblock \bibinfo{title}{Modeling of the eclipsing binaries in the globular
  cluster ngc 5466}.
\newblock \bibinfo{journal}{AJ} \bibinfo{volume}{389}, \bibinfo{pages}{590}.
\bibitem[{Kaluzny et~al.(2013)Kaluzny, Thompson, Rozyczka, Dotter, Krzeminski,
  Pych, Rucinski, Burley and Shectman}]{kaluzny2013cluster}
\bibinfo{author}{Kaluzny, J.}, \bibinfo{author}{Thompson, I.},
  \bibinfo{author}{Rozyczka, M.}, \bibinfo{author}{Dotter, A.},
  \bibinfo{author}{Krzeminski, W.}, \bibinfo{author}{Pych, W.},
  \bibinfo{author}{Rucinski, S.}, \bibinfo{author}{Burley, G.},
  \bibinfo{author}{Shectman, S.}, \bibinfo{year}{2013}.
\newblock \bibinfo{title}{The cluster ages experiment (case). v. analysis of
  three eclipsing binaries in the globular cluster m4}.
\newblock \bibinfo{journal}{The Astronomical Journal} \bibinfo{volume}{145},
  \bibinfo{pages}{43}.
\bibitem[{{Kaluzny} et~al.(1997){Kaluzny}, {Thompson} and
  {Krzeminski}}]{kaluzny1997ccd}
\bibinfo{author}{{Kaluzny}, J.}, \bibinfo{author}{{Thompson}, I.B.},
  \bibinfo{author}{{Krzeminski}, W.}, \bibinfo{year}{1997}.
\newblock \bibinfo{title}{Ccd photometry of faint variable stars in the field
  of the globular cluster m4}.
\newblock \bibinfo{journal}{Astronomical Journal} \bibinfo{volume}{113},
  \bibinfo{pages}{2219--2225}.
\newblock \DOIprefix\doi{10.1086/118432}.
\bibitem[{{Kaluzny} et~al.(2013){Kaluzny}, {Thompson}, {Rozyczka} and
  {Krzeminski}}]{Kaluzny2013new}
\bibinfo{author}{{Kaluzny}, J.}, \bibinfo{author}{{Thompson}, I.B.},
  \bibinfo{author}{{Rozyczka}, M.}, \bibinfo{author}{{Krzeminski}, W.},
  \bibinfo{year}{2013}.
\newblock \bibinfo{title}{{The Clusters AgeS Experiment (CASE): Variable Stars
  in the Globular Cluster M4}}.
\newblock \bibinfo{journal}{Acta Astronomica} \bibinfo{volume}{63},
  \bibinfo{pages}{181--201}.
\bibitem[{Kanatas et~al.(1995)Kanatas, Griffiths, Dickens and
  Penny}]{kanatas1995ccd}
\bibinfo{author}{Kanatas, I.}, \bibinfo{author}{Griffiths, W.},
  \bibinfo{author}{Dickens, R.}, \bibinfo{author}{Penny, A.},
  \bibinfo{year}{1995}.
\newblock \bibinfo{title}{Ccd photometry in the globular cluster m4}.
\newblock \bibinfo{journal}{MNRAS} \bibinfo{volume}{272}, \bibinfo{pages}{265}.
\bibitem[{Kiron et~al.(2011)Kiron, Sriram and Rao}]{kiron2011photometric}
\bibinfo{author}{Kiron, Y.R.}, \bibinfo{author}{Sriram, K.},
  \bibinfo{author}{Rao, P.V.}, \bibinfo{year}{2011}.
\newblock \bibinfo{title}{Photometric parameters, distance and period-colour
  study of contact binary stars in the globular cluster $\omega$ centauri}.
\newblock \bibinfo{journal}{Bulletin of the Astronomical Society of India}
  \bibinfo{volume}{39}.
\bibitem[{Kiron et~al.(2012)Kiron, Sriram and
  Vivekananda~Rao}]{kiron2012photometric}
\bibinfo{author}{Kiron, Y.R.}, \bibinfo{author}{Sriram, K.},
  \bibinfo{author}{Vivekananda~Rao, P.}, \bibinfo{year}{2012}.
\newblock \bibinfo{title}{A photometric study of contact binaries v3 and v4 in
  ngc 2539}.
\newblock \bibinfo{journal}{Bulletin of the Astronomical Society of India, Vol.
  40, p. 51} \bibinfo{volume}{40}, \bibinfo{pages}{51}.
\bibitem[{Ko{\c{c}}ak et~al.(2020)Ko{\c{c}}ak, {\.I}{\c{c}}li and
  Yakut}]{koccak2020photometric}
\bibinfo{author}{Ko{\c{c}}ak, D.}, \bibinfo{author}{{\.I}{\c{c}}li, T.},
  \bibinfo{author}{Yakut, K.}, \bibinfo{year}{2020}.
\newblock \bibinfo{title}{Photometric study of close binary stars in the m35,
  m67, and m71 galactic clusters}.
\newblock \bibinfo{journal}{arXiv preprint arXiv:2002.05159} .
\bibitem[{Kopacki and Pigulski(1995)}]{kopacki1995uw}
\bibinfo{author}{Kopacki, G.}, \bibinfo{author}{Pigulski, A.},
  \bibinfo{year}{1995}.
\newblock \bibinfo{title}{Uw cvn--an eclipsing binary system of w uma type}.
\newblock \bibinfo{journal}{Acta Astronomica} \bibinfo{volume}{45},
  \bibinfo{pages}{753}.
\bibitem[{Kouzuma(2019)}]{kouzuma2019starspots}
\bibinfo{author}{Kouzuma, S.}, \bibinfo{year}{2019}.
\newblock \bibinfo{title}{Starspots in contact and semi-detached binary
  systems}.
\newblock \bibinfo{journal}{Publications of the Astronomical Society of Japan}
  \bibinfo{volume}{71}, \bibinfo{pages}{21}.
\bibitem[{Lanza and Rodon{\`o}(2002)}]{lanza2002gravitational}
\bibinfo{author}{Lanza, A.}, \bibinfo{author}{Rodon{\`o}, M.},
  \bibinfo{year}{2002}.
\newblock \bibinfo{title}{Gravitational quadrupole-moment variations in active
  binaries}.
\newblock \bibinfo{journal}{Astronomische Nachrichten} \bibinfo{volume}{323},
  \bibinfo{pages}{424--431}.
\bibitem[{Li(2018)}]{li2018photometric}
\bibinfo{author}{Li, K.}, \bibinfo{year}{2018}.
\newblock \bibinfo{title}{Photometric study of the eclipsing blue straggler
  v205 in the globular cluster ngc 5139}.
\newblock \bibinfo{journal}{New Astronomy} \bibinfo{volume}{59},
  \bibinfo{pages}{60--64}.
\bibitem[{Li et~al.(2017)Li, Hu, Chen and Guo}]{li2017comp}
\bibinfo{author}{Li, K.}, \bibinfo{author}{Hu, S.}, \bibinfo{author}{Chen, X.},
  \bibinfo{author}{Guo, D.}, \bibinfo{year}{2017}.
\newblock \bibinfo{title}{Comprehensive photometric study of an extremely low
  mass ratio deep contact binary in the globular cluster m 4}.
\newblock \bibinfo{journal}{Publications of the Astronomical Society of Japan}
  \bibinfo{volume}{69}, \bibinfo{pages}{79}.
\bibitem[{Li and Qian(2013a)}]{li2013two}
\bibinfo{author}{Li, K.}, \bibinfo{author}{Qian, S.B.}, \bibinfo{year}{2013}a.
\newblock \bibinfo{title}{Two contact binaries at different evolutionary stages
  in the globular cluster ngc 6397}.
\newblock \bibinfo{journal}{New Astronomy} \bibinfo{volume}{25},
  \bibinfo{pages}{12--18}.
\bibitem[{Li and Qian(2013b)}]{li2013one}
\bibinfo{author}{Li, K.}, \bibinfo{author}{Qian, S.B.}, \bibinfo{year}{2013}b.
\newblock \bibinfo{title}{Two unusual contact binaries in the globular cluster
  m54}.
\newblock \bibinfo{journal}{New Astronomy} \bibinfo{volume}{22},
  \bibinfo{pages}{57--61}.
\bibitem[{Li et~al.(2021)Li, Xia, Kim, Gao, Hu, Guo, Gao, Chen and
  Guo}]{li2021photometric}
\bibinfo{author}{Li, K.}, \bibinfo{author}{Xia, Q.Q.}, \bibinfo{author}{Kim,
  C.H.}, \bibinfo{author}{Gao, X.}, \bibinfo{author}{Hu, S.M.},
  \bibinfo{author}{Guo, D.F.}, \bibinfo{author}{Gao, D.Y.},
  \bibinfo{author}{Chen, X.}, \bibinfo{author}{Guo, Y.N.},
  \bibinfo{year}{2021}.
\newblock \bibinfo{title}{Photometric study and absolute parameter estimation
  of six totally eclipsing contact binaries}.
\newblock \bibinfo{journal}{The Astronomical Journal} \bibinfo{volume}{162},
  \bibinfo{pages}{13}.
\bibitem[{Liu et~al.(2011)Liu, Qian and Fern{\'a}ndez-Laj{\'u}s}]{liu2011two}
\bibinfo{author}{Liu, L.}, \bibinfo{author}{Qian, S.B.},
  \bibinfo{author}{Fern{\'a}ndez-Laj{\'u}s, E.}, \bibinfo{year}{2011}.
\newblock \bibinfo{title}{Two extreme contact binary systems in the nearest
  globular cluster m4}.
\newblock \bibinfo{journal}{Monthly Notices of the Royal Astronomical Society}
  \bibinfo{volume}{415}, \bibinfo{pages}{1509--1522}.
\bibitem[{Liu et~al.(2014)Liu, Qian and Laj{\'u}s}]{liu2014identification}
\bibinfo{author}{Liu, L.}, \bibinfo{author}{Qian, S.B.},
  \bibinfo{author}{Laj{\'u}s, E.F.}, \bibinfo{year}{2014}.
\newblock \bibinfo{title}{Identification of a deep contact binary in the field
  of the globular cluster 47 tuc}.
\newblock \bibinfo{journal}{New Astronomy} \bibinfo{volume}{26},
  \bibinfo{pages}{116}.
\bibitem[{Lucy(1967)}]{lucy1967structure}
\bibinfo{author}{Lucy, L.}, \bibinfo{year}{1967}.
\newblock \bibinfo{title}{Structure of w ursae majoris stars}.
\newblock \bibinfo{journal}{The Astronomical Journal} \bibinfo{volume}{72},
  \bibinfo{pages}{309--310}.
\bibitem[{Ma et~al.(2022)Ma, Liu, Zhang, Hu and L{\"u}}]{ma2022photometric}
\bibinfo{author}{Ma, S.}, \bibinfo{author}{Liu, J.Z.}, \bibinfo{author}{Zhang,
  Y.}, \bibinfo{author}{Hu, Q.}, \bibinfo{author}{L{\"u}, G.L.},
  \bibinfo{year}{2022}.
\newblock \bibinfo{title}{A photometric study of two contact binaries: Crts
  j025408. 1+ 265957 and crts j012111. 1+ 272933}.
\newblock \bibinfo{journal}{Research in Astronomy and Astrophysics}
  \bibinfo{volume}{22}, \bibinfo{pages}{095017}.
\bibitem[{Mateo(1996)}]{mateo1996photometric}
\bibinfo{author}{Mateo, M.}, \bibinfo{year}{1996}.
\newblock \bibinfo{title}{Photometric binary stars in globular clusters}, in:
  \bibinfo{booktitle}{the Origins, Evolution, and Destinies of Binary Stars in
  Clusters}, p.~\bibinfo{pages}{21}.
\bibitem[{Milone et~al.(2014)Milone, Marino, Bedin, Piotto, Cassisi, Dieball,
  Anderson, Jerjen, Asplund, Bellini et~al.}]{milone2014m}
\bibinfo{author}{Milone, A.}, \bibinfo{author}{Marino, A.},
  \bibinfo{author}{Bedin, L.}, \bibinfo{author}{Piotto, G.},
  \bibinfo{author}{Cassisi, S.}, \bibinfo{author}{Dieball, A.},
  \bibinfo{author}{Anderson, J.}, \bibinfo{author}{Jerjen, H.},
  \bibinfo{author}{Asplund, M.}, \bibinfo{author}{Bellini, A.}, et~al.,
  \bibinfo{year}{2014}.
\newblock \bibinfo{title}{The m 4 core project with hst--ii. multiple stellar
  populations at the bottom of the main sequence}.
\newblock \bibinfo{journal}{MNRAS} \bibinfo{volume}{439},
  \bibinfo{pages}{1588}.
\bibitem[{Milone et~al.(2012)Milone, Piotto, Bedin, Aparicio, Anderson,
  Sarajedini, Marino, Moretti, Davies, Chaboyer et~al.}]{milone2012acs}
\bibinfo{author}{Milone, A.}, \bibinfo{author}{Piotto, G.},
  \bibinfo{author}{Bedin, L.}, \bibinfo{author}{Aparicio, A.},
  \bibinfo{author}{Anderson, J.}, \bibinfo{author}{Sarajedini, A.},
  \bibinfo{author}{Marino, A.}, \bibinfo{author}{Moretti, A.},
  \bibinfo{author}{Davies, M.B.}, \bibinfo{author}{Chaboyer, B.}, et~al.,
  \bibinfo{year}{2012}.
\newblock \bibinfo{title}{The acs survey of galactic globular clusters-xii.
  photometric binaries along the main sequence}.
\newblock \bibinfo{journal}{A\&A} \bibinfo{volume}{540}, \bibinfo{pages}{A16}.
\bibitem[{Milone et~al.(2008)Milone, Piotto, Bedin and
  Sarajedini}]{milone2008photometric}
\bibinfo{author}{Milone, A.}, \bibinfo{author}{Piotto, G.},
  \bibinfo{author}{Bedin, L.}, \bibinfo{author}{Sarajedini, A.},
  \bibinfo{year}{2008}.
\newblock \bibinfo{title}{Photometric binaries in 50 globular clusters}.
\newblock \bibinfo{journal}{MmSAI} \bibinfo{volume}{79}, \bibinfo{pages}{623}.
\bibitem[{Mochejska et~al.(2002)Mochejska, Kaluzny, Thompson and
  Pych}]{mochejska2002clusters}
\bibinfo{author}{Mochejska, B.}, \bibinfo{author}{Kaluzny, J.},
  \bibinfo{author}{Thompson, I.}, \bibinfo{author}{Pych, W.},
  \bibinfo{year}{2002}.
\newblock \bibinfo{title}{Clusters ages experiment: Hot subdwarfs and luminous
  white dwarf candidates in the field of the globular cluster m4}.
\newblock \bibinfo{journal}{AJ} \bibinfo{volume}{124}, \bibinfo{pages}{1486}.
\bibitem[{Mochnacki and Doughty(1972)}]{mochnacki1972model}
\bibinfo{author}{Mochnacki, S.}, \bibinfo{author}{Doughty, N.},
  \bibinfo{year}{1972}.
\newblock \bibinfo{title}{A model for the totally eclipsing w ursae majoris
  system aw uma}.
\newblock \bibinfo{journal}{Monthly Notices of the Royal Astronomical Society}
  \bibinfo{volume}{156}, \bibinfo{pages}{51--65}.
\bibitem[{Nascimbeni et~al.(2014)Nascimbeni, Bedin, Heggie, van~den Berg,
  Giersz, Piotto, Brogaard, Bellini, Milone, Rich et~al.}]{nascimbeni2014m}
\bibinfo{author}{Nascimbeni, V.}, \bibinfo{author}{Bedin, L.},
  \bibinfo{author}{Heggie, D.}, \bibinfo{author}{van~den Berg, M.},
  \bibinfo{author}{Giersz, M.}, \bibinfo{author}{Piotto, G.},
  \bibinfo{author}{Brogaard, K.}, \bibinfo{author}{Bellini, A.},
  \bibinfo{author}{Milone, A.}, \bibinfo{author}{Rich, R.}, et~al.,
  \bibinfo{year}{2014}.
\newblock \bibinfo{title}{The m 4 core project with hst--iii. search for
  variable stars in the primary field}.
\newblock \bibinfo{journal}{MNRAS} \bibinfo{volume}{442},
  \bibinfo{pages}{2381}.
\bibitem[{O'Connell(1951)}]{o1951so}
\bibinfo{author}{O'Connell, D.}, \bibinfo{year}{1951}.
\newblock \bibinfo{title}{The so-called periastron effect in close eclipsing
  binaries; new variable stars (fifth list)}.
\newblock \bibinfo{journal}{Riverview College Observatory publications; v. 2,
  no. 6= whole no. 10; Riverview College Observatory publications; v. 2, no.
  6., Riverview, NSW:[Riverview College Observatory, 1951], p. 85-100: ill.; 27
  cm.} \bibinfo{volume}{2}, \bibinfo{pages}{85--100}.
\bibitem[{Pecaut and Mamajek(2013)}]{pecaut2013intrinsic}
\bibinfo{author}{Pecaut, M.J.}, \bibinfo{author}{Mamajek, E.E.},
  \bibinfo{year}{2013}.
\newblock \bibinfo{title}{Intrinsic colors, temperatures, and bolometric
  corrections of pre-main-sequence stars}.
\newblock \bibinfo{journal}{The Astrophysical Journal Supplement Series}
  \bibinfo{volume}{208}, \bibinfo{pages}{9}.
\bibitem[{Popper and Ulrich(1977)}]{popper1977evolutionary}
\bibinfo{author}{Popper, D.M.}, \bibinfo{author}{Ulrich, R.K.},
  \bibinfo{year}{1977}.
\newblock \bibinfo{title}{The evolutionary status of rs canum venaticorum
  binaries.}
\newblock \bibinfo{journal}{The Astrophysical Journal} \bibinfo{volume}{212},
  \bibinfo{pages}{L131--L134}.
\bibitem[{Pothuneni et~al.(2023)Pothuneni, Devarapalli and
  Jagirdar}]{pothuneni2023first}
\bibinfo{author}{Pothuneni, R.R.}, \bibinfo{author}{Devarapalli, S.P.},
  \bibinfo{author}{Jagirdar, R.}, \bibinfo{year}{2023}.
\newblock \bibinfo{title}{The first photometric and spectroscopic study of
  contact binary v2840 cygni}.
\newblock \bibinfo{journal}{Research in Astronomy and Astrophysics}
  \bibinfo{volume}{23}, \bibinfo{pages}{025017}.
\bibitem[{Priya et~al.(2013)Priya, Sriram and Rao}]{priya2013photometric}
\bibinfo{author}{Priya, D.S.}, \bibinfo{author}{Sriram, K.},
  \bibinfo{author}{Rao, P.V.}, \bibinfo{year}{2013}.
\newblock \bibinfo{title}{Photometric study of an eclipsing binary in
  praesepe}.
\newblock \bibinfo{journal}{Research in Astronomy and Astrophysics}
  \bibinfo{volume}{13}, \bibinfo{pages}{465}.
\bibitem[{Qian(2001)}]{qian2001possible}
\bibinfo{author}{Qian, S.}, \bibinfo{year}{2001}.
\newblock \bibinfo{title}{A possible relation between the period change and the
  mass ratio for w-type contact binaries}.
\newblock \bibinfo{journal}{MNRAS} \bibinfo{volume}{328}, \bibinfo{pages}{635}.
\bibitem[{Qian(2003)}]{qian2003overcontact}
\bibinfo{author}{Qian, S.}, \bibinfo{year}{2003}.
\newblock \bibinfo{title}{Are overcontact binaries undergoing thermal
  relaxation oscillation with variable angular momentum loss?}
\newblock \bibinfo{journal}{MNRAS} \bibinfo{volume}{342},
  \bibinfo{pages}{1260}.
\bibitem[{Richer et~al.(2004)Richer, Brewer, Fahlman, Kalirai, Stetson, Hansen,
  Rich, Ibata, Gibson and Shara}]{richer2004concerning}
\bibinfo{author}{Richer, H.B.}, \bibinfo{author}{Brewer, J.},
  \bibinfo{author}{Fahlman, G.}, \bibinfo{author}{Kalirai, J.},
  \bibinfo{author}{Stetson, P.}, \bibinfo{author}{Hansen, B.},
  \bibinfo{author}{Rich, R.}, \bibinfo{author}{Ibata, R.},
  \bibinfo{author}{Gibson, B.}, \bibinfo{author}{Shara, M.},
  \bibinfo{year}{2004}.
\newblock \bibinfo{title}{Concerning the white dwarf cooling age of m4: A
  response to the paper by de marchi et al. on" a different interpretation of
  recent hst observations"}.
\newblock \bibinfo{journal}{arXiv preprint astro-ph/0401446} .
\bibitem[{Rucinski(1969)}]{rucinski1969proximity}
\bibinfo{author}{Rucinski, S.}, \bibinfo{year}{1969}.
\newblock \bibinfo{title}{The proximity effects in close binary systems. ii.
  the bolometric reflection effect for stars with deep convective envelopes}.
\newblock \bibinfo{journal}{Acta Astronomica} \bibinfo{volume}{19},
  \bibinfo{pages}{245}.
\bibitem[{Rucinski(2000)}]{rucinski2000w}
\bibinfo{author}{Rucinski, S.M.}, \bibinfo{year}{2000}.
\newblock \bibinfo{title}{W uma-type binary stars in globular clusters}.
\newblock \bibinfo{journal}{The Astronomical Journal} \bibinfo{volume}{120},
  \bibinfo{pages}{319}.
\bibitem[{Rucinski and Duerbeck(1997)}]{rucinski1997absolute}
\bibinfo{author}{Rucinski, S.M.}, \bibinfo{author}{Duerbeck, H.W.},
  \bibinfo{year}{1997}.
\newblock \bibinfo{title}{Absolute magnitude calibration for the w uma-type
  systems based on hipparcos data}.
\newblock \bibinfo{journal}{PASP} \bibinfo{volume}{109}, \bibinfo{pages}{1340}.
\bibitem[{Rukmini et~al.(2005)Rukmini, Rao and Sriram}]{rukmini2005photometric}
\bibinfo{author}{Rukmini, J.}, \bibinfo{author}{Rao, P.V.},
  \bibinfo{author}{Sriram, K.}, \bibinfo{year}{2005}.
\newblock \bibinfo{title}{Photometric study of w uma type binary in the old
  cluster ngc 6791}.
\newblock \bibinfo{journal}{Astrophysics and Space Science}
  \bibinfo{volume}{299}, \bibinfo{pages}{109--116}.
\bibitem[{Saad(2005)}]{saad2005light}
\bibinfo{author}{Saad, S.}, \bibinfo{year}{2005}.
\newblock \bibinfo{title}{Light curve analysis of two new w uma stars in m15}.
\newblock \bibinfo{journal}{Astrophysics and Space Science}
  \bibinfo{volume}{296}, \bibinfo{pages}{301--304}.
\bibitem[{Safonova et~al.(2016)Safonova, Mkrtichian, Hasan, Sutaria, Brosch,
  Gorbikov and Joseph}]{safonova2016search}
\bibinfo{author}{Safonova, M.}, \bibinfo{author}{Mkrtichian, D.},
  \bibinfo{author}{Hasan, P.}, \bibinfo{author}{Sutaria, F.},
  \bibinfo{author}{Brosch, N.}, \bibinfo{author}{Gorbikov, E.},
  \bibinfo{author}{Joseph, P.}, \bibinfo{year}{2016}.
\newblock \bibinfo{title}{Search for low-mass objects in the globular cluster
  m4. i. detection of variable stars}.
\newblock \bibinfo{journal}{The Astronomical Journal} \bibinfo{volume}{151},
  \bibinfo{pages}{27}.
\bibitem[{Scargle(1982)}]{scargle1982studies}
\bibinfo{author}{Scargle, J.D.}, \bibinfo{year}{1982}.
\newblock \bibinfo{title}{Studies in astronomical time series analysis.
  ii-statistical aspects of spectral analysis of unevenly spaced data}.
\newblock \bibinfo{journal}{ApJ} \bibinfo{volume}{263},
  \bibinfo{pages}{835--853}.
\bibitem[{Stetson et~al.(2014)Stetson, Braga, Dall’Ora, Bono, Buonanno,
  Ferraro, Iannicola, Marengo and Neeley}]{stetson2014optical}
\bibinfo{author}{Stetson, P.}, \bibinfo{author}{Braga, V.},
  \bibinfo{author}{Dall’Ora, M.}, \bibinfo{author}{Bono, G.},
  \bibinfo{author}{Buonanno, R.}, \bibinfo{author}{Ferraro, I.},
  \bibinfo{author}{Iannicola, G.}, \bibinfo{author}{Marengo, M.},
  \bibinfo{author}{Neeley, J.}, \bibinfo{year}{2014}.
\newblock \bibinfo{title}{Optical and near-infrared ubvrijhk photometry for the
  rr lyrae stars in the nearby globular cluster m4 (ngc 6121) 1}.
\newblock \bibinfo{journal}{PASP} \bibinfo{volume}{126}, \bibinfo{pages}{521}.
\bibitem[{Van~Hamme(1993)}]{van1993new}
\bibinfo{author}{Van~Hamme, W.}, \bibinfo{year}{1993}.
\newblock \bibinfo{title}{New limb-darkening coefficients for modeling binary
  star light curves}.
\newblock \bibinfo{journal}{Astronomical Journal (ISSN 0004-6256), vol. 106,
  no. 5, p. 2096-2117} \bibinfo{volume}{106}, \bibinfo{pages}{2096--2117}.
\bibitem[{Van~Hamme and Wilson(2003)}]{van2003stellar}
\bibinfo{author}{Van~Hamme, W.}, \bibinfo{author}{Wilson, R.},
  \bibinfo{year}{2003}.
\newblock \bibinfo{title}{Stellar atmospheres in eclipsing binary models}, in:
  \bibinfo{booktitle}{GAIA Spectroscopy: Science and Technology}, p.
  \bibinfo{pages}{323}.
\bibitem[{Wallace(2018)}]{Wallace2018}
\bibinfo{author}{Wallace, J.J.}, \bibinfo{year}{2018}.
\newblock \bibinfo{title}{M4 membership catalog from gaia proper motions}.
\newblock \bibinfo{journal}{Research Notes of the AAS} \bibinfo{volume}{2},
  \bibinfo{pages}{213}.
\newblock \URLprefix \url{https://dx.doi.org/10.3847/2515-5172/aaf1a2},
  \DOIprefix\doi{10.3847/2515-5172/aaf1a2}.
\bibitem[{Wilsey and Beaky(2009)}]{wilsey2009revisiting}
\bibinfo{author}{Wilsey, N.J.}, \bibinfo{author}{Beaky, M.M.},
  \bibinfo{year}{2009}.
\newblock \bibinfo{title}{Revisiting the o'connell effect in eclipsing binary
  systems}, in: \bibinfo{booktitle}{The Society for Astronomical Sciences 28th
  Annual Symposium on Telescope Science. Held May 19-21, 2009 at Big Bear Lake,
  CA. Published by the Society for Astronomical Sciences., p. 107}, p.
  \bibinfo{pages}{107}.
\bibitem[{Wilson and Devinney(1971)}]{wilson1971realization}
\bibinfo{author}{Wilson, R.E.}, \bibinfo{author}{Devinney, E.J.},
  \bibinfo{year}{1971}.
\newblock \bibinfo{title}{Realization of accurate close-binary light curves:
  application to mr cygni}.
\newblock \bibinfo{journal}{ApJ} \bibinfo{volume}{166}, \bibinfo{pages}{605}.
\bibitem[{Yakut and Eggleton(2005)}]{yakut2005evolution}
\bibinfo{author}{Yakut, K.}, \bibinfo{author}{Eggleton, P.P.},
  \bibinfo{year}{2005}.
\newblock \bibinfo{title}{Evolution of close binary systems}.
\newblock \bibinfo{journal}{ApJ} \bibinfo{volume}{629}, \bibinfo{pages}{1055}.
\bibitem[{Zechmeister and K{\"u}rster(2009)}]{zechmeister2009generalised}
\bibinfo{author}{Zechmeister, M.}, \bibinfo{author}{K{\"u}rster, M.},
  \bibinfo{year}{2009}.
\newblock \bibinfo{title}{The generalised lomb-scargle periodogram-a new
  formalism for the floating-mean and keplerian periodograms}.
\newblock \bibinfo{journal}{A \& A} \bibinfo{volume}{496},
  \bibinfo{pages}{577}.
\bibitem[{Zloczewski et~al.(2013)Zloczewski, Kaluzny, Rozyczka, Krzeminski,
  Mazur and Thompson}]{zloczewski2013proper}
\bibinfo{author}{Zloczewski, K.}, \bibinfo{author}{Kaluzny, J.},
  \bibinfo{author}{Rozyczka, M.}, \bibinfo{author}{Krzeminski, W.},
  \bibinfo{author}{Mazur, B.}, \bibinfo{author}{Thompson, I.},
  \bibinfo{year}{2013}.
\newblock \bibinfo{title}{A proper motion study of the globular clusters m4,
  m12, m22, ngc 3201, ngc 6362 and ngc 6752}.
\newblock \bibinfo{journal}{arXiv preprint arXiv:1301.1198} .

\end{thebibliography}
	
	
	
	

	
	
	

	
	\bio{}
	\endbio
	
	\bio{}
	\endbio
	
\end{document}